\documentclass[aps, prb, twocolumn, showpacs]{revtex4-1}

\usepackage{bm}
\usepackage{calc}
\usepackage{float}
\usepackage{color}
\usepackage{graphicx}
\usepackage{mathtools}
\usepackage{amsmath, amsfonts}

\def\abs#1{\left \vert #1 \right \vert}
\def\etal{\textit{et al.\ }}

\DeclarePairedDelimiterX{\norm}[1]{\lVert}{\rVert}{#1}

\begin{document}

\title{Peculiarities in the gravitational field of a filamentary ring}

\author{D{\'a}niel Schumayer}
\email{daniel.schumayer@otago.ac.nz}
\author{David A. W. Hutchinson}
\affiliation{The Dodd-Walls Centre for Photonic and Quantum Technologies,
             Department of Physics, University of Otago, Dunedin 9106,
             New Zealand}

\date{\today}

\begin{abstract}
    The gravitational field of a massive, filamentary ring is considered.
    We provide an analytic expression for the gravitational potential and
    demonstrate that the exact gravitational potential and its gradient, thus
    the gravitational force-field, is not central. Hence it is a good candidate
    to discuss the difference between the concepts of center of mass and center
    of gravity. We focus on other consequences of reduced symmetry, e.g., only
    the $z$-component of the angular momentum is conserved. However, the remnant
    high symmetry of this system also ensures that there are special classes of
    motions which are restricted to invariant subspaces, thus, depending on the
    initial condition, the dynamics of a point particle is integrable. We also
    show that periodic orbits in the equatorial plane external to the ring are
    possible, but only if the angular momentum is above a threshold value. In
    this case the orbits are stable.
\end{abstract}

\maketitle

\section{Introduction}

From the earliest evidence regarding the activities of our ancestors,
we know that stargazing and attempting to explain celestial events, the motion
of Sun, Moon, and, perhaps, constellations were important to everyday life. In
cave drawings and rock carvings one can find motifs of archaeo-astronomical
significance.\cite{Mourao2009, Norris2011}

Why all these celestial bodies move the way they appear to move? What forces of
nature govern their behaviour? These questions occupied natural philosophers for
centuries and, with numerous contributors, they received a quantitative answer
in Newton's {\emph{Philosophi{\ae} Naturalis Principia Mathematica}} (Mathematical
Principles of Natural Philosophy) in 1687. Apart from giving the foundational
rules for classical mechanics, Newton also provided a theory of gravity. His
inverse-square law could explain and predict --to a higher accuracy than its
contemporary theories-- how planets are orbiting in our solar system.

In most undergraduate courses on classical mechanics this Newtonian theory of
gravitation is taught, predominantly applied to only point particles and/or
spherical objects, such as the Sun and Earth.\cite{Jones1993, Halliday2004,
Giancoli2008, Serway2000, Urone2012, Hewitt2013, Knight2013} In the latter case,
it is undoubtedly mentioned that one may calculate the gravitational attraction
between two spherical objects if their masses were concentrated in their
individual centers and one rather calculated the force between these point
particles. Furthermore, the same process applies if density of each of these
spherical objects varies only with the distance from the center. Planets and
stars can be treated as such spherical objects, at least in the leading order.
Despite the necessarily simplistic analytic treatment of the dynamics of
gravitating bodies in undergraduate entry courses, one can find useful
supplementary material\cite{Green2015} with numerical simulations which could
help students develop a good intuition about motion under the influence of
non-central forces.

The idea of replacing an extensive object with an abstract massive point
particle is extremely powerful, conceptually simple, and essential to efficient
modelling of physical processes. It is also rarely mentioned that for
non-spherical objects one may not apply this rule of thumb directly. Seldom are
the concepts of ``center of mass'' and ``center of gravity'' distinguished.
\footnote{In all textbooks cited before all mentions that in uniform
gravitational field the center of mass and center of gravity coincide. While
this practice is defensible from the view of practicality, we feel that
pedagogically is questionable: an exceptional case (point particle or perfectly
spherical objects) is taught. Only Ref. \cite{Serway2000}, among the already
cited textbooks, treats non-uniform mass distribution and non-spherical bodies
explicitly in a subsection.}
The former is uniquely determined by the shape and mass distribution of
a body, thus it is a well-defined point. However, the center of gravity depends
on the interaction of two gravitating bodies, thus it may not be a fixed point
as the case of spherical bodies would indicate. The example system examined
below demonstrates the inapplicability of this rule of thumb for a very simple
system.
\begin{figure}[b!]
    \begin{center}
    \includegraphics[width=85mm]{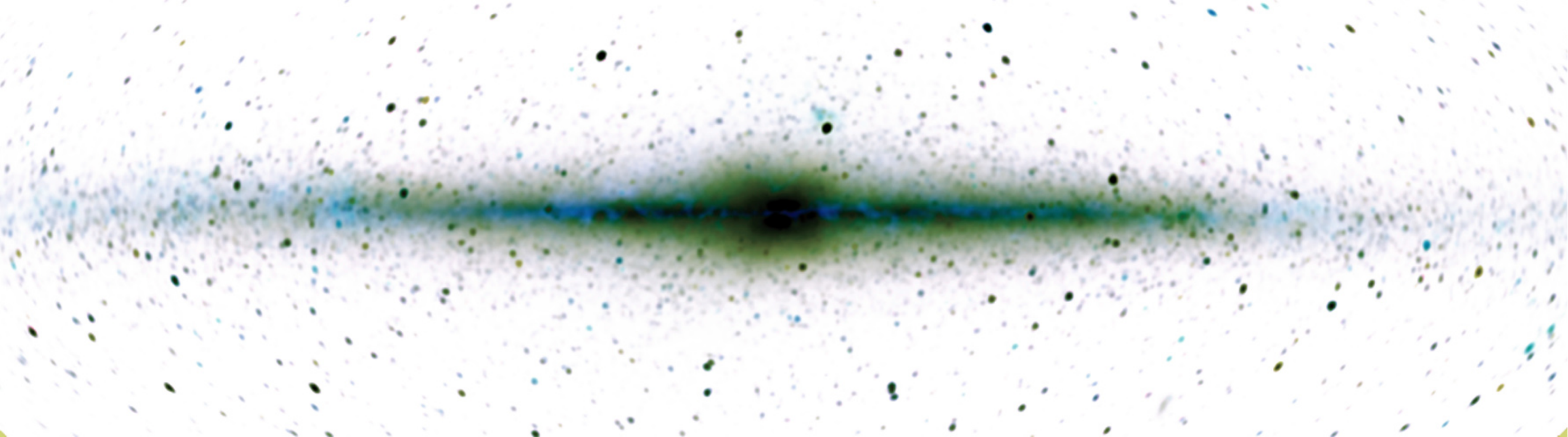}
    \caption{\label{fig:MilkyWayFromSide}
             (Color online) Inverted false-color image of the near-infrared sky
             as assembled using 1.25 (blue), 2.2 (green), and 3.5\,$\mu$m (red)
             wavelengths in the {\textsc{cobe}} project. The plane of the
             Milky Way is horizontal across the middle and the Galactic center
             is located at the center of the image. The image shows both the
             thin disk and central bulge populations of stars in our galaxy. The
             diameter of the thin disk is approximately 50,000\,pc\footnote{
             Parsec, abbreviated as pc, is an astronomical unit of length and it
             is equal to about 3.261 light-years.}, its width is 350\,pc, while
             the width of the central bulge is circa 3500\,pc. (Courtesy of
             {\textsc{nasa/cobe}})
             }
    \end{center}
\end{figure}
Here we study the gravitational potential of a massive, ideal ring and the
motion of a massive point particle in this field. The simplicity, finitude and
the high degree of symmetry of this geometry motivated our analysis. One might
object, saying that an infinite line would truly have even simpler geometry.
However, with uniform mass distribution, it would have infinite mass. Also we
would like to preserve as much of the spherical symmetry as possible, and the
field of a ring has cylindrical symmetry.

There are two other --admittedly less weighty-- motivations for this work.
First, the orbits of numerous planets can be considered to be ring-like, and
during their revolution around the center their own gravitational field perturb
the motion of other planets, moons, comets, etc. Gauss averaging theorem
states~\cite{Gauss1818, Abad1997, Boccaletti2013} that the secular perturbative
effect of a revolving object can be calculated by spreading its mass around its
orbit uniformly and determining the perturbation of this ring on the target
object. Thus the gravitational field of a uniform ring does indeed have
significance in astrophysics.

\begin{figure}[b!]
    \begin{center}
    \includegraphics[angle=-90, width=85mm]
                    {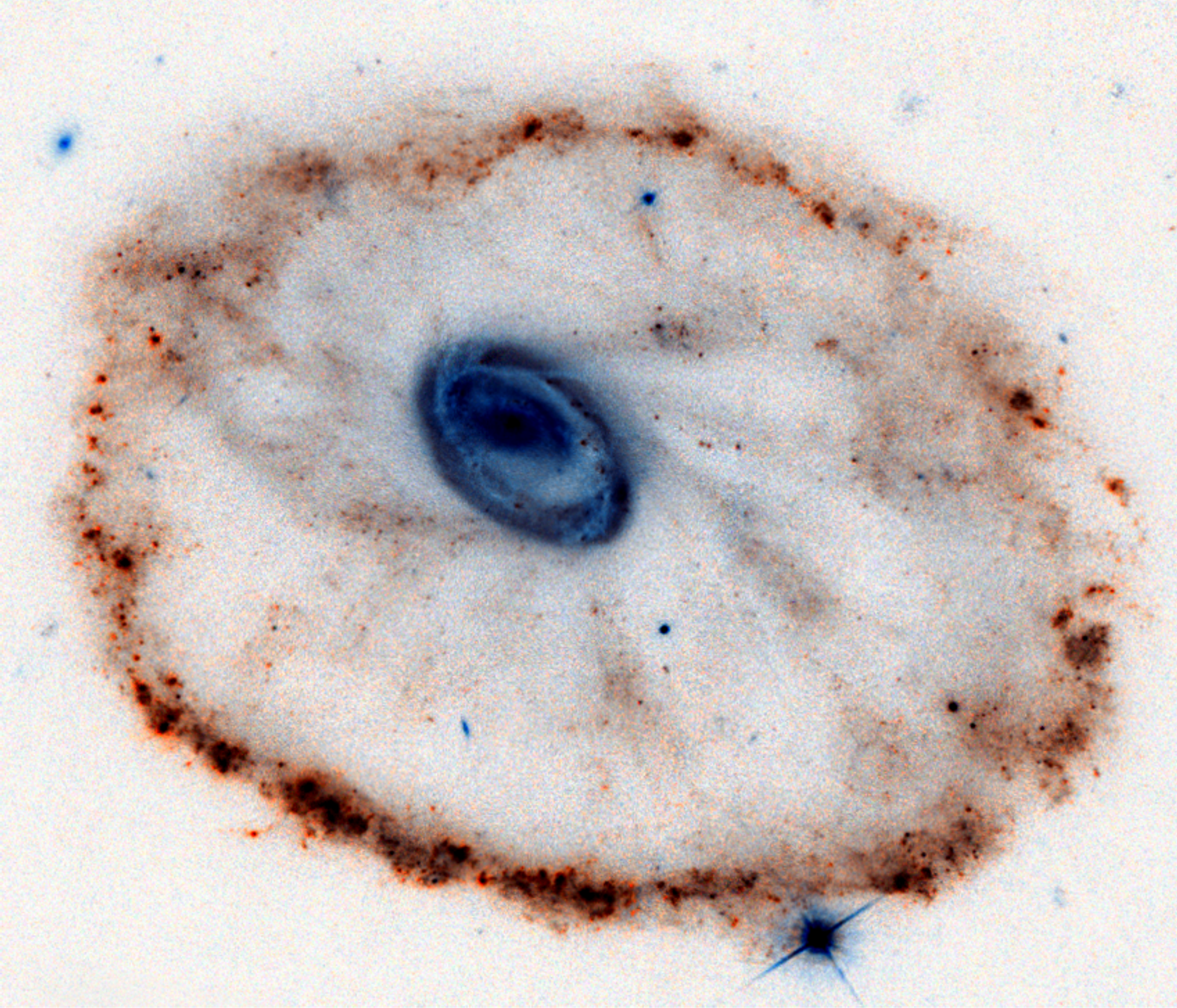}
    \caption{\label{fig:potw1036a_CartwheelGalaxy}
             (Color online) Image of the Cartwheel Galaxy ({\textsc{eso}}
             350-40) taken by the Hubble Space Telescope (courtesy of ESA/Hubble
             \& NASA). The color scheme is inverted: the deep sky is white and
             the bright stars are dark. This galaxy is located about 500 million
             light-years away in the Sculptor constellation. Its shape resembles
             a wagon wheel and developed after a smaller galaxy collided with
             and passed through the middle of a larger galaxy about 200 million
             years ago. The collision produced shock waves sweeping gas and dust
             radially outwards as a ripple. The outermost ring is approximately
             1.5 times the size of our Milky Way.
             }
    \end{center}
\end{figure}

Second, ring structures have actual importance in galactic dynamics as well.
Several celestial objects are more or less spherical, e.g., stars, planets,
black holes, but there are other objects, e.g., galaxies, interstellar dust and
clouds, whose geometry varies strongly. In our own galaxy, in the Milky Way,
there are billions of stars revolving around the center in a thin disk, see
Figure~\ref{fig:MilkyWayFromSide}. The thickness of this imaginary disk is
approximately hundred and fifty times smaller than its radius, but a substantial
part of the galaxy's mass is in the disk~\cite{Zeilik2002, Carroll2017}.
Similarly there are ring galaxies in the Universe, where the millions of stars
constituting the galaxy are located along a ring with no or small amount of
luminous matter visible in their interiors,\cite{Theys1976, Theys1977} e.g.,
Cartwheel Galaxy in Figure~\ref{fig:potw1036a_CartwheelGalaxy}.

In this work we aim to provide a simple enough system which can be analytically
analyzed at the graduate level, but whose gravitational field holds peculiar
features, e.g., it produces a non-central force field. We believe the
investigation of such system is beneficial for students as in most cases they
only see spherically symmetric bodies whose exterior gravitational field can be
determined by that of an imaginary point particle. While the analysis presented
here is relatively simple, it may stimulate the Reader's appreciation of the
longstanding challenge of calculating the gravitational potential/field of
celestial bodies. As an excellent example, we mention the landing of a small
robot Phil\ae\ on the comet 67P/Churyumov-Gerasimenko traveling at tens of
thousands of kilometers per second. That was the first occasion in history that
an expedition successfully landed on a comet. The estimation of the
gravitational field was crucial to achieve such a landing and later measurements
allowed for the determination of the inner structure of that
comet.\cite{Paetzold2016} Furthermore we may also point at the resurgent
interest in analyzing gravitational fields of extended objects, mainly that of
the Earth~\cite{Hofmeister2018}, due to the commercial interest. There are a
handful companies, e.g., SpaceX, RocketLab, Blue Origin, manufacturing and
launching rockets and placing satellites in orbits on a regular basis. In order
to plan the orbits of these spacecrafts and monitor their trajectories, more and
more precise models and planning will be required.

\section{General formalism in a nutshell}

Newton's law of gravitation states that the force between two massive point
particles is proportional to the masses of interacting particles, inversely
proportional to the square of their distances and lies in the direction of the
line connecting the particles. The proportionality constant is denoted by $G$
in the following.

Since the gravitational field is conservative, the force can be deduced from a
scalar potential, $V(\bm{r})$, satisfying Poisson's equation with mass
distribution, $\rho(\bm{r})$. Often the density is given and one seeks the
gravitational potential. One can express the potential knowing the Green's
function of the adjoint boundary value problem. Since the Laplace operator
equipped with Cauchy boundary condition at infinity (the potential must vanish
infinitely far from the gravitating body) is self-adjoint, the adjoint Green's
function, $h(\bm{r} \vert \bm{\xi})$, coincides with the Green's function of the
original problem, $g(\bm{r} \vert \bm{\xi})$. Thus the Green's function,
$g(\bm{r} \vert \bm{\xi})= \frac{1}{4\pi} \norm{\bm{r}-\boldsymbol{\xi}}^{-1}$,
provides the potential
\begin{equation} \label{eq:PotentialFromGreensFunction}
    V(\mathbf{r}, \rho)
    =
    - G \!\int{\frac{\rho(\bm{\xi})}{\norm{\bm{r} - \bm{\xi}}} \,d\xi}.
\end{equation}
Although the natural domains of $\bm{r}$ and $\bm{\xi}$ are identical, the
three-dimensional space, we only analyze cases where the density distribution,
$\rho$, is restricted to a domain of finite size, $\Omega$, which is occupied by
the fixed mass distribution. For an idealized, filamentary ring $\Omega$ denotes
the circumference of the ring, while in the case of an annulus and a disk
$\Omega$ is the surface area of these objects. Since $\rho$ is assumed to vanish
completely outside of $\Omega$, the integral on the right hand side of
Eq.~\eqref{eq:PotentialFromGreensFunction} is also interpreted to be over only
$\Omega$.

It is worth recognizing how the potential is affected by scaling the density and
distance. If $\lambda$ and $\eta$ denote two non-zero numbers, then
Eq.~\eqref{eq:PotentialFromGreensFunction} shows that scaling the density by
a factor of $\lambda$ would result in scaling the potential by $\lambda$ too,
i.e., $V(\bm{r}, \lambda \rho) = \lambda V(\bm{r}, \rho)$, and similarly
$V(\eta \bm{r}, \rho) = \frac{1}{\eta} V(\bm{r}, \rho)$. As a consequence, one
could opt for calculating the gravitational potential of a ring of unit mass and
unit radius, since the result can be scaled to the actual mass and radius.
This scaling property of the potential can be useful when one writes a numerical
code, for example, to simulate the motion of an object under the influence of
the ring's gravitational potential. However, for the sake of clarity and being
explicit we keep both the mass and the radius of the ring as variables.

Here we mention two recent analyses of the electrostatic potential and
electric-field of a charged ring. As the gravitational potential and the
electrostatic potential both satisfy Poisson's equation, the same mathematical
machinery can be employed in both cases. However, the dynamics of an external
charge heavily depends on whether the interaction is attractive or negative.
Zypman considered \cite{Zypman2006} the electric field of a uniformly charged
ring and put the emphasis on an intuitive introduction of the potential and
visualizing the electric field using a mathematical software package. In
connection to our interests here, the sketch of the electric field in the $[xz]$
plane implicitly showed that the field, and thus the force on a point charge is
not central. Selvaggi \etal employed \cite{Selvaggi2007} a different approach
and demonstrated --without detailed introduction of the toroidal functions--
that the electric potential can be expressed in a compact form using the Fourier
cosine expansion of the $\norm{\bm{r}- \bm{\xi}}^{-1}$, in which expansion the
toroidal functions appear as coefficients of the $\cos(m\varphi)$ functions.
This approach has some numerically attractive features: any non-uniform charge
distribution can be systematically analyzed after calculating the Fourier
expansion of the charge distribution on the ring, and for each term in this
expansion there is a single corresponding toroidal function. However, the choice
of toroidal functions of first or second kind may require further thoughts. In
\citet{Selvaggi2007} only the second kind is used, while in the mathematical
literature it is used predominantly for the interior problem, $r_{\!\perp} < R$.

\section{Gravitational potential of a ring}

Let us examine an ideal ring whose total mass, $M$, is distributed uniformly
along the circumference of a circle of radius $R$, thus $2 \pi R \rho = M$. This
system is regularly in the focus of classes on classical mechanics and
non-relativistic electrodynamics, although the gravitational or electrostatic
potential is determined almost always along the symmetry axis, $z$,
perpendicular to the plane of the ring.

The plane of the ring divides the three-dimensional space into two halves and
the potential will thus have a reflection symmetry with respect to this plane,
i.e., $V(x, y, z) = V(x, y, -z)$. Furthermore, the potential is rotationally
invariant along the $z$-axis which crosses the center of the ring and
perpendicular to the plane of the ring.

One may determine the gravitational potential at point $P$ by carrying out the
integration in Eq.~\eqref{eq:PotentialFromGreensFunction} along the ring. Due to
all symmetries, however, one may fix the Cartesian coordinate system in such a
way that $P$ has zero $y$ coordinate.
\begin{figure}[h!]
    \begin{center}
    \includegraphics[width=80mm]{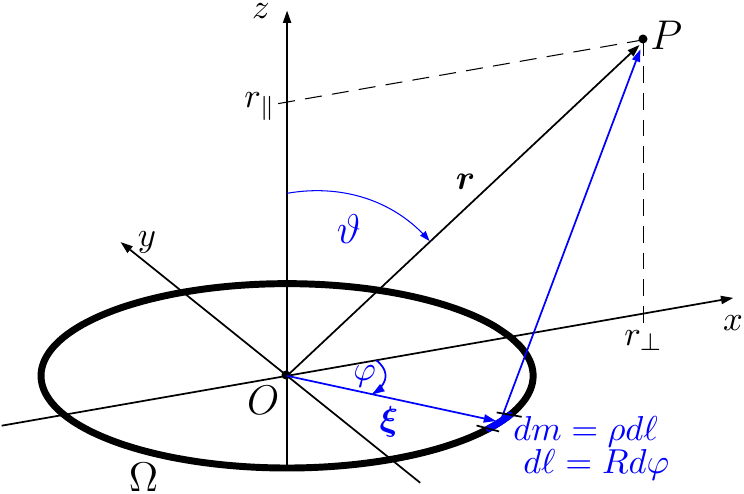}
    \caption{\label{fig:Ring_Schematic}
             (Color online) The geometry and denotation used for determining the
             gravitational potential of a ring, $\Omega$, with radius $R$ and
             homogeneous mass distribution, $\rho$.
             }
    \end{center}
\end{figure}

The gravitational potential of the ideal ring at point $\bm{r} = (x, 0, z)$
is given
\begin{align}
    V(\bm{r})
    &=
    - G \int_{\Omega}{\frac{\rho}{\norm{\bm{r} - \bm{\xi}}} \,d\xi}
    \nonumber
    \\
    &=
    - G
    \int_{0}^{2\pi}
        {\!\frac{\rho R d\varphi}
              {\sqrt{ (r_{\!\perp}\! + R)^2 + z^{2} -
                      2 r_{\!\perp} R (1 + \cos{\!(\varphi)})
                    }
              }
        }
    \nonumber
    \\
    &=
    - \frac{2 G M}{\pi p}
    \int_{0}^{\frac{\pi}{2}}
        {\!\frac{d\varphi}
              {\sqrt{1 - k^{2} \cos^{2}{\!(\varphi)}}}
        },
\end{align}
where we have introduced $\chi^{2} = (r_{\!\perp}\! + R)^{2} + z^{2}$, which is
the distance of point $P$ from the furthest point of the ring, and $k^{2}= 4 R\,
r_{\!\perp}/\chi^{2}$. In the standard nomenclature of the field of elliptic
integrals, parameter $k$ is called the modulus.~\cite{Byrd1954, Abramowitz1972}

The integral above cannot be given in elementary functions, and itself defines a
new special function, the complete elliptic integral of the first
kind,\cite{Byrd1954} $K(k)$, with modulus $k$. It is apparent that this integral
is real only if $k \le 1$, and it is singular only if $k=1$. One may easily
check that $k$ remains in the $[0, 1]$ interval for any $\bm{r}$ and it becomes
unity only at the points of the ring itself. Consequently the gravitational
potential becomes singular only ``on'' the ring. This is expected and it is due
to the inherent nature of Newton's gravitational law itself. However, the
strength of singularity is milder than $-1/\norm{\bm{r}}$ as anticipated from
Newton's law for point particles. Since $\cos^{2}{\!(\varphi)} \sim 1 -
\varphi^{2}$ for $\abs{\varphi} \ll 1$, thus for $k=1$ the denominator of the
integrand is $\sqrt{1 - k^{2} \cos^{2}{\!(\varphi) }} \sim \varphi$. Therefore
the integral locally behaves as $\ln{\!(\varphi)}$, i.e., the singularity is
logarithmic rather than power-law.

\begin{figure}[h!]
    \begin{center}
    \includegraphics[width=85mm]{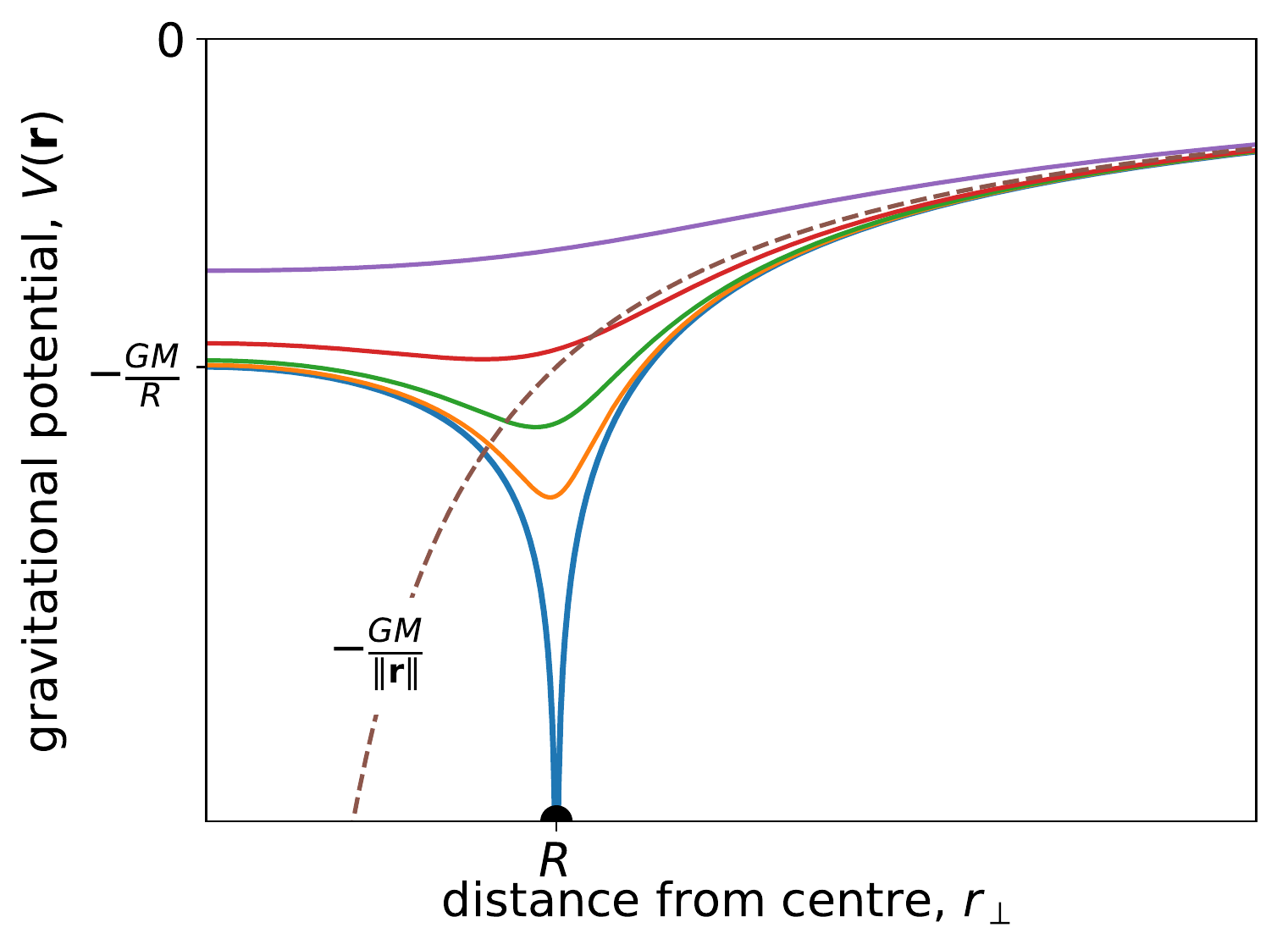}
    \caption{\label{fig:Ring_PotentialAtFixedHeight}
             Gravitational potential $V(\bm{r})$, for fixed heights $z$, as a
             function of the distance, $r_{\!\perp}$ from the center of the ring.
             The singular curve correspond to the potential in the plane of the
             ring ($z=0$), while the height is finite and positive for the other
             curves. The black blob at $r_{\!\perp}=R$ represents the ring.
             }
    \end{center}
\end{figure}

Finally the gravitational potential is\cite{Lass1983}
\begin{equation}
    \label{eq:Ring_Potential}
    V(\bm{r}) = - \frac{2 G M}{\pi \chi} K(k),
\end{equation}
and it is plotted as a function of $r_{\!\perp}$ for some fixed height values in
Figure~\ref{fig:Ring_PotentialAtFixedHeight}. The singular curve corresponds to
the gravitational potential in the plane of the ring, i.e., $z=0$, while the for
other curves, $z \ne 0$. The potential along the symmetry axis has a value of
$-GM/\sqrt{R^{2} + z^{2}}$, in agreement with standard elementary treatment.
While Figure~\ref{fig:Ring_PotentialAtFixedHeight} clearly shows that the
potential indeed becomes shallower at the center for higher $z$ value, it is
interesting to note that the gradient of $V(\bm{r})$ for small $z$ values is
negative, but as $z$ increases it turns to positive. In other words, the
potential is repulsive at the center of the ring if the height is below some
threshold height, $z^{\ast}$, but for heights $z^{\ast} \le z$ the center
becomes attractive and it is the global minimum of the potential. It is
fascinating that in a simple gravitational system the potential changes its
qualitative behavior along a symmetry axis not smoothly but abruptly.

We can determine the threshold height at which the potential sways from
attractive to repulsive. We have noticed that at fixed $z$ the potential $V(\bm{r})$
develops a minimum at a finite $r_{\!\perp}$. The location of this minimum,
i.e., the $(r_{\!\perp}, z)$ can be determined by taking the first derivative of
$V(\bm{r})$ with respect to $r_{\!\perp}$ and equating it to zero. After some
straightforward calculation one obtains the equation, which establishes the
relationship between $r_{\!\perp}$ and $z$ at the minimum:
\begin{equation}
    \label{eq:DeterminingTheLocationOfMinima}
    2 (1-k^{2}) (E(k) - K(k)) + k^{2} (1-r_{\!\perp}) E(k) = 0,
\end{equation}
where $E(k)$ is the complete elliptic integral of the second
kind.\cite{Byrd1954} Although this equation seems intractable to express
$r_{\!\perp}$ in terms of $z$ at the minimum or vice versa, let us remember that
at the threshold height, $z^{\ast}$, the symmetry axis becomes the global
minimum, thus at this height this equation must support a solution of $(r_{\!
\perp}, z) = (0, z^{\ast})$. Furthermore, at $r_{\!\perp} = 0$ the modulus
becomes zero, and both complete elliptic integrals are not only finite for such
modulus, but $K(0) = E(0)$. However, we cannot blindly substitute $r_{\!\perp} =
0$ and $k=0$ into the equation above, as it would lead to the trivial identity;
the first term vanishes because of the factor $(E(k)-K(k))$, while the second
factor is zero due to $k^{2}$. Thus we have to determine the solution of
Eq.~\eqref{eq:DeterminingTheLocationOfMinima} in the limit of $k \rightarrow 0$.
The series expansion of both special functions are well known (see formulas
900.00 and 900.07 in Byrd~\&~Friedman's monograph\cite{Byrd1954}) and can be
readily employed in this equation. After inserting these expansions, truncated
at the $k^{2}$ order, we arrive at
\begin{equation}
    r_{\!\perp}
    \left \lbrack
        1- \frac{3}{2} \frac{R^{2}}{(R+r_{\!\perp})^{2} + z^{2}}
    \right \rbrack = 0.
\end{equation}
The $r_{\!\perp}=0$ is always a solution for any $z$ value. However this
expression does not say whether that is the minimum or maximum of the potential.
From Figure~\ref{fig:Ring_PotentialAtFixedHeight} we deduced that for $z <
z^{\ast}$ the potential will attain its local maximum at $r_{\!\perp}=0$. As we
are seeking the location of minima we need to examine the expression in the
square brackets, and perhaps express $z$ in terms of $r_{\!\perp}$:
\begin{equation}
    z
    =
    \pm R \sqrt{\frac{3}{2} - \left ( 1 + \frac{r_{\!\perp}}{R} \right)^{2}}.
\end{equation}
As we derived this expression in the $k \rightarrow 0$ limit, this result is
valid only around the symmetry axis. We can immediately read off $z^{\ast}$, as
$r_{\!\perp} \rightarrow 0^{+}$. Thus $z^{\ast} = \pm R/\sqrt{2}$. The Reader
may verify this result as follows. As
Figure~\ref{fig:Ring_PotentialAtFixedHeight} indicates the center of the ring is
always an extremum of the potential; at $z=0$ the potential has a maximum, while
at higher altitude along the $z$-axis become a global minimum of the potential
develops. In other words the second derivative of the potential, with respect to
$z$, changes sign from positive to negative at the threshold point, $z^{\ast}$.
Thus one needs to solve the equation
\begin{equation*}
    \frac{\partial^{2} V(\textbf{r})}{\partial z^{2}}
    =
    \frac{2 GM}{\pi} \frac{1}{\chi^{4}}
    \Bigl \lbrack (R^{2} - 3z^{2}) E(k) + z^{2} K(k) \Bigr \rbrack
    = 0
\end{equation*}
for $z$ with fixed $r_{\perp}=0$ thus $k=0$. Although this expression still
contains both elliptic integrals,\footnote{An even simpler approach would be to
first substitute $r_{\perp} = 0$ into the expression of the gravitational
potential and obtain $V(z)=-GM/\sqrt{R^{2} + z^{2}}$ and then differentiate this
expression twice and solve the $\partial_{z}^{2}V(z)=0$ equation for $z^{\ast}$.}
their values can be factored out after substituting $k=0$ in, since $E(0)=K(0)$.
One then arrives at the second order equation $R^{2}-2z^{2}=0$, which indeed
confirms the value of $z^{\ast}$.
\begin{figure}[bht!]
    \begin{center}
    \includegraphics[width=80mm]{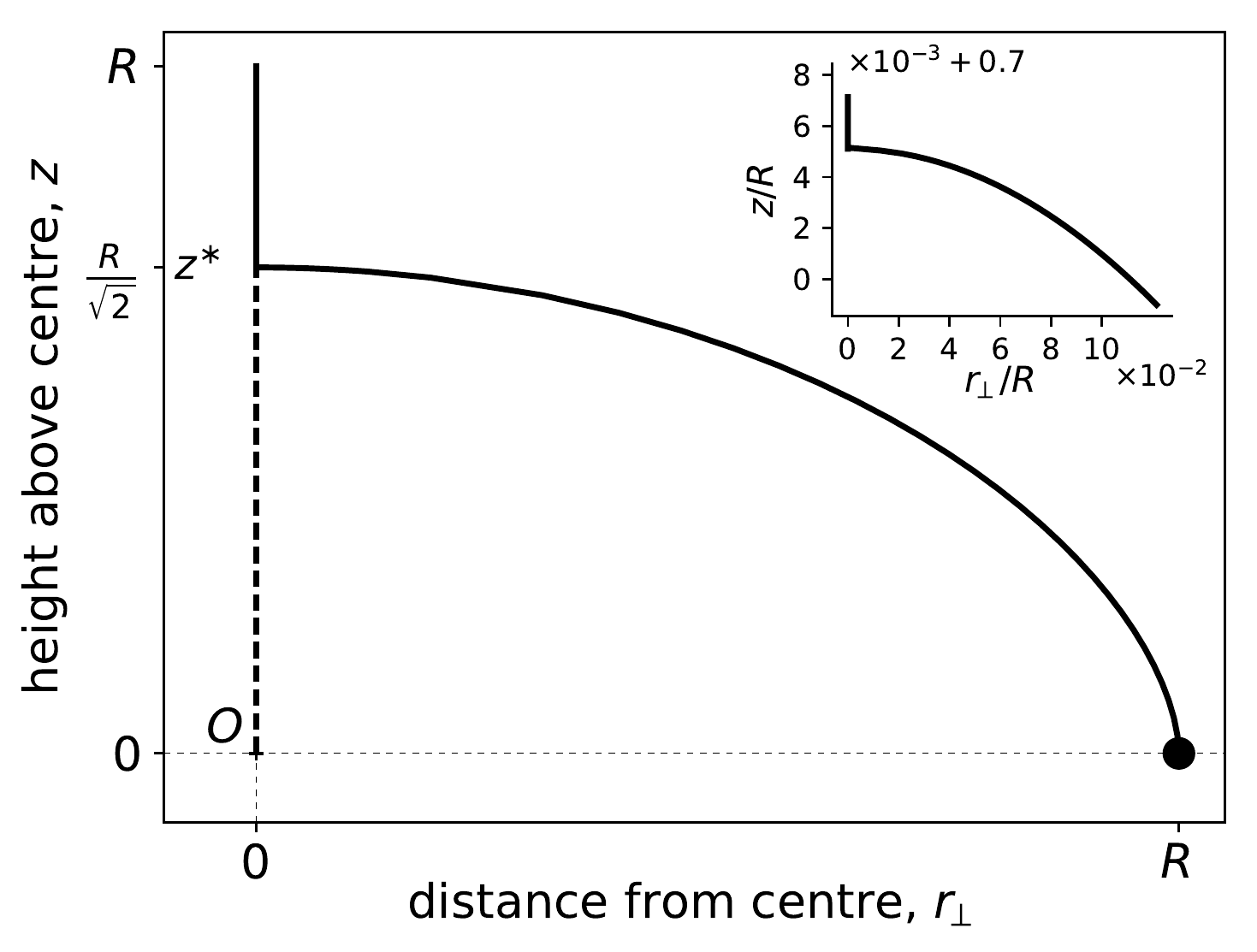}
    \caption{\label{fig:Ring_LocationOfMinima}
             The gravitational potential has a global minimum ``inside'' the
             ring. The curve shows the location of the minimum for heights, $z
             \le R$. It is interesting to note that such minimum exists only for
             $0 < \abs{z} \le R/\sqrt{2}$, otherwise the only minimum is at the
             center of the ring. The thick solid line corresponds to the
             location of minima, while the thick dashed line at the symmetry
             axis represents the local maxima.
             }
    \end{center}
\end{figure}

Another interesting feature of the gravitational potential in
Figure~\ref{fig:Ring_PotentialAtFixedHeight} is the appearance of a global
minimum provided $z$ is kept fixed and $z \ne 0$, within the ring, $0 \le
r_{\!\perp} \le R$. This fact may --incorrectly-- suggest that a particle can be
in stable equilibrium in the inner vicinity of the ring, away from the center,
since no force would be exerted on a particle as $\partial_{r_{\!\perp}}V$
vanishes at the minimum. The crux of this fallacy lies in omitting the gradient
of the potential in the $z$ direction at the minimum, i.e., $\bm{F} \propto
-\nabla V = (-\partial_{r_{\!\perp}}\!V) \bm{e}_{r_{\!\perp}} \! + (-\partial_{
z} V) \bm{e}_{z} = (-\partial_{z} V) \bm{e}_{z} \ne \bm{0}$.

\begin{figure}[b!]
    \begin{center}
    \includegraphics[width=85mm]{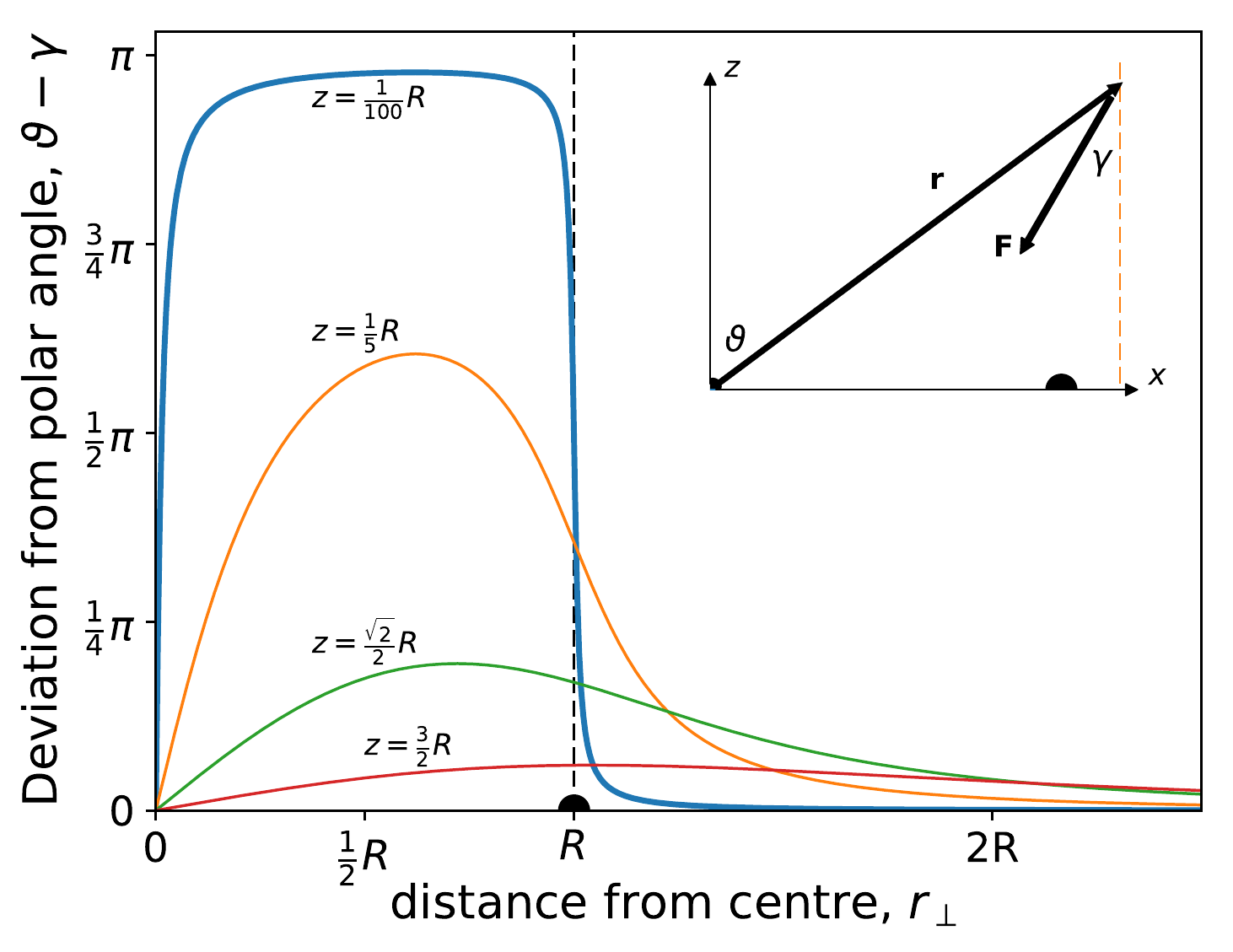}
    \caption{\label{fig:Ring_AngleDeviation}
             (Color online) The angle difference, $\vartheta-\gamma$, is plotted
             at four, fixed heights, $z/R=\frac{1}{100}$, $\frac{1}{5}$, $\frac{
             \sqrt{2}}{2}$, and $\frac{3}{2}$ as a function of $r_{\!\perp}$.
             The inset shows the geometric definition of these angles;
             $\vartheta$ is the standard polar angle in the spherical coordinate
             system, and $\gamma$ is the angle of the gravitational force, $\bm{
             F}$, relative to the vertical. For central force fields $\vartheta$
             and $\gamma$ are either identical or opposite of each other, thus
             their difference is zero or $\pi$. It is apparent, however, that
             for finite $z$ their difference varies with $r_{\!\perp}$ and $z$,
             thus the gravitational field of the ring is not a central force
             field.
            }
    \end{center}
\end{figure}

Before we shift our focus to the dynamics in the gravitational field of a ring,
let us briefly examine and quantify how much the force field deviates from a
central field. The cylindrical symmetry guarantees that neither the potential,
nor the force depend on the azimuthal angle, $\varphi$. However, after
expressing $\chi$ and $k$ (see Eq.~\eqref{eq:Ring_Potential}) in spherical
coordinates
\begin{equation*}
    \chi^{2} = R^{2} + r^{2} + 2Rr \sin{\!(\vartheta)}
    \quad \mbox{and}\quad
    k =
    \frac{4R r \sin{\!(\vartheta)}}{\chi^{2}},
\end{equation*}
it becomes apparent that the force does depend on $\vartheta$ and, as a
consequence, it cannot represent a central force. In order to visualize and
quantify how much the force deviates from a central field, let us calculate the
difference of angles, $\gamma = \angle(\bm{F}, -\bm{e}_{z})$ and the polar
angle, $\vartheta$ (see inset of Figure~\ref{fig:Ring_AngleDeviation}). For
radial force these two angles should be either equal in absolute values, thus
their difference is either zero or $\pi$. In Fig.~\ref{fig:Ring_AngleDeviation}
we have plotted their difference, $\vartheta - \gamma$, as a function of
$r_{\!\perp}$ for fixed heights, $z$. The graphs show that this difference
changes abruptly closer to the plane of the ring, and confirms that for $z=0$
the two domains (inside and outside the ring) behave quite differently, although
in both cases the force points towards the ring. Inside the ring $\gamma =
-\frac{\pi}{ 2}$, while outside $\gamma = \frac{\pi}{2}$.

As $z$ increases the angle difference drops, but still changes significantly,
reaches its maximum ``inside'' the ring, and diminishes rapidly as $R <
r_{\!\perp}$. However, as $z$ increases further the angle difference vanishes
and the maximum becomes very wide.

\section{Dynamics in the gravitational field of a ring}

In classical mechanics one often examines the dynamics of a particle under the
influence of a central force; the force can be written as $\bm{F}(\bm{r}) =
F(\bm{r}) \bm{e}_{r}$, i.e., the magnitude of the force, $F(\bm{r})$, and its
direction, $\bm{e}_{r}$, are decoupled and the direction is always parallel to
$\bm{e}_{r}$. Since physically important forces, e.g., gravitational attraction
and electrostatic interaction, fall into this category the analysis of this
special type of forces is warranted. It is proven that the trajectory of a
particle moving in a central force field lies in a plane, and the angular
momentum of the particle, $\bm{L}$, is a conserved quantity. If, furthermore,
$F(\bm{r})$ depends only on $\norm{\bm{r}}$, then the force field is
conservative as well, thus total mechanical energy is conserved along any
trajectory. This special case is the one which is often analyzed in university
courses.

However, even the gravitational attraction of an ideal ring is more complicated
than this special central, conservative force field. While $V(\bm{r})$ is
cylindrically symmetric, it is not spherically symmetric, ergo $F(\bm{r}) =
F(\norm{r}, \vartheta)$ depends on the distance from the center, $\norm{\bm{r}}$,
and on the polar angle, $\vartheta$, too. There are only two special sets of
trajectories along which a particle would experience a central, conservative
force field: (a) along the $z$-axis, and (b) on any trajectory lying in the
$[xy]$-plane of the ring. Therefore in these two cases we can invoke energy and
angular momentum conservation, and thereby reduce the degrees of freedom to one.
Since these two sets are also part of the invariant subspaces of the potential,
a particle remains in these subspaces provided its initial momentum is also
parallel to these subspaces: in case (a) $\bm{p} \propto \bm{e}_{z}$ and (b)
$\bm{p} \perp \bm{e}_{z}$. In case (a) this constraint also means that the
angular momentum vanishes identically.

For later use it seems worthwhile to record the expression of force at an
arbitrary point in space. For sake of transparency the force is rescaled by
a unit $f_{0}$, and also the coordinates are given in units of $R$. Due to
rotational symmetry around the $z$-axis, the force at any point $P$ should lie
in the plane which includes $P$ and the entire $z$-axis, i.e., the force depends
only on the distance of $P$ from the $z$-axis and from the equatorial plane.
Therefore the force can be expressed as
\begin{equation}
    \bm{F} =
    -\frac{x}{r_{\!\perp}}
    \!\left ( \frac{\partial U}{\partial r_{\!\perp}} \!\right ) \bm{e}_{x}
    -\frac{y}{r_{\!\perp}}
    \!\left ( \frac{\partial U}{\partial r_{\!\perp}} \!\right ) \bm{e}_{y} -
    \left ( \frac{\partial U}{\partial z} \!\right ) \bm{e}_{z}
\end{equation}
where $U = mV$ is the interaction potential between the ring and the massive
particle orbiting in the field of the ring. After some relatively
straightforward calculation we can express the force in the form:
\begin{equation}
    \label{eq:GeneralMotionInTheFieldOfRing}
    \bm{F}
    =
    \frac{2}{\pi} \frac{G M m}{\chi^{3}} \frac{1}{1-k^{2}}
    \bigl ( \Upsilon x, \Upsilon y,  E(k) z \bigr )
\end{equation}
with the dimensionless factor
\begin{equation}
    \Upsilon
    =
    \frac{R^{2} + z^{2} - r^{2}_{\!\perp}}{2 R^{2}} \bigl ( E(k)-K(k) \bigr ) +
    \frac{R - r_{\!\perp}}{R} K(k).
\end{equation}
In general, the component of the total angular momentum, $\bm{L}$, along an axis
about which the field is symmetrical is always conserved,\cite{Landau1982} thus
$L_{z}$ is conserved. Based on the results shown above we can easily check for
this by calculating the Poisson-bracket of $H$ and $L_{z}$
\begin{equation}
   \left \lbrace H\,; L_{z}\!\right \rbrace
    =
    \sum_{i=1}^{3}
        {\!\left \lbrack
            \frac{\partial H}{\partial q_{i}}
            \frac{\partial L_{z}}{\partial p_{i}}
            -
            \frac{\partial H}{\partial p_{i}}
            \frac{\partial L_{z}}{\partial q_{i}}
         \right \rbrack}
\end{equation}
where $\bm{q}=(q_{1}, q_{2}, q_{3}) = (x, y, z)$ and analogously $\bm{p} =
(p_{1}, p_{2}, p_{3}) = (p_{x}, p_{y}, p_{z})$.
This Poisson-bracket involves the partial derivatives with respect to
the coordinates and the conjugate momenta. Since the Hamiltonian of a massive
particle moving in the gravitational potential of a ring has the traditional
form $H(\bm{q}, \bm{p}) =\frac{1}{2m} \,\bm{p}^{2} + U(\bm{q})$, i.e., the
coordinates and the momenta are separated, the derivatives of $H$ are simple:
$\partial H/\partial q_{i} = \partial U/\partial q_{i} = -F_{i}$ while
$\partial H/\partial p_{i} = \frac{1}{m} p_{i}$. The derivatives of $L_{z}$ are
also easy to calculate from $L_{z}= q_{1} p_{2} - q_{2}p_{1}$, hence
\begin{equation}
   \left \lbrace H\,; L_{z}\!\right \rbrace
    =
    F_{x} y - F_{y} x
    =
    -\frac{1}{r_{\!\perp}}
    \left ( \!\frac{\partial U}{\partial r_{\!\perp}} \!\right ) (xy- yx)
    = 0.
\end{equation}
Furthermore, cylindrical symmetry or, in other words, the conservation of
$L_{z}$, allows one to reduce the number of degrees of freedom of the generic
Hamiltonian from three ($x$, $y$, and $z$) to two ($r_{\!\perp}$ and $z$) and
obtain
\begin{equation}
    H (r_{\!\perp}, z)
    =
    \frac{1}{2m} \!\left ( p_{\!\perp}^{2} + p_{z}^{2} \right ) +
    \frac{L_{z}^{2}}{2 m r_{\!\perp}^{2}} + U(r_{\!\perp}, z).
\end{equation}

\subsection{Motion along the $z$ axis}

\begin{figure}[h!]
    \begin{center}
    \includegraphics[width=80mm]{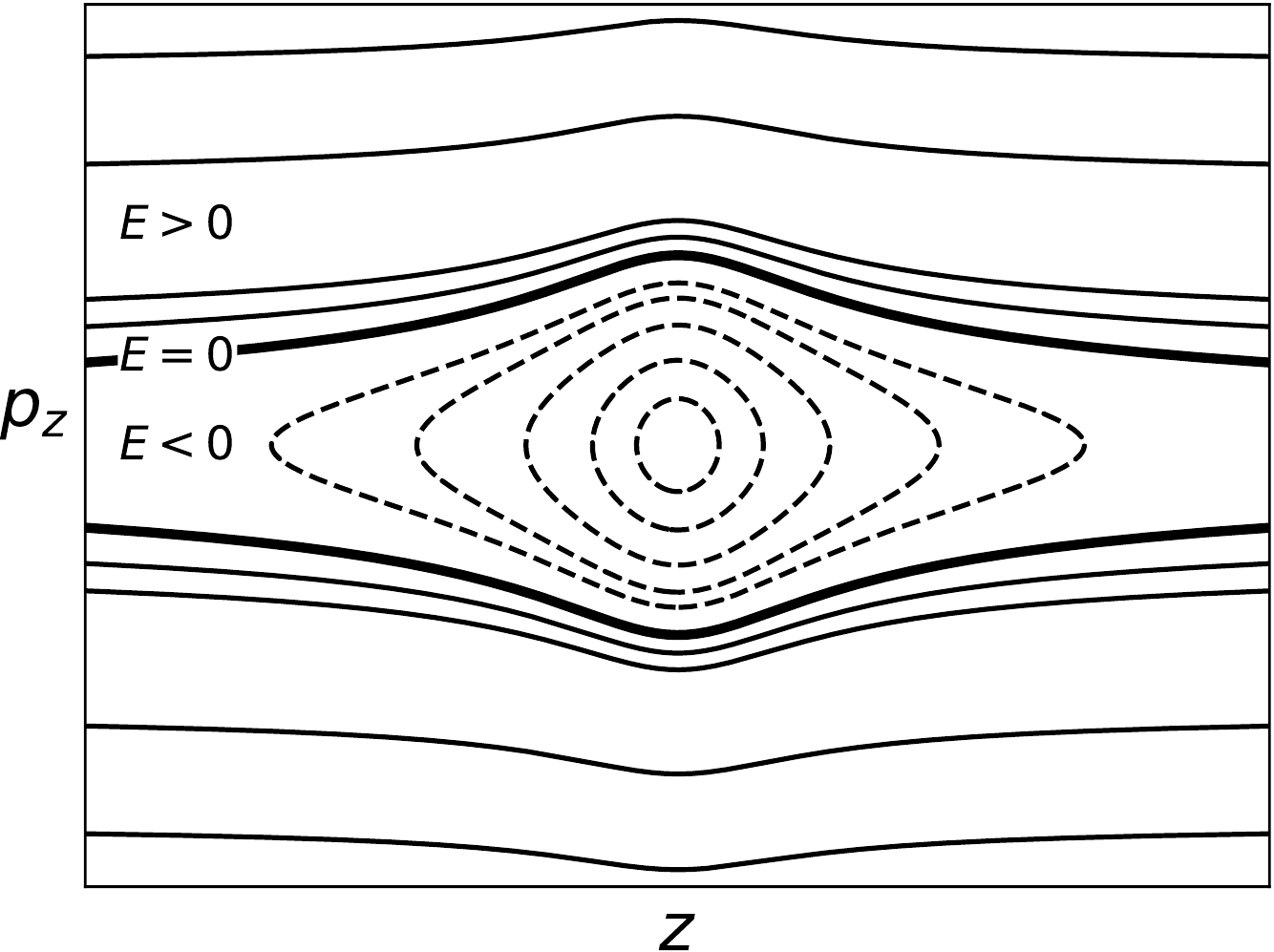}
    \caption{\label{fig:Ring_PhasePlot_MotionAlongZAxis}
             Contour plot of the total mechanical energy, $H(z, p_{z})$, over
             the phase space for a particle moving along the $z$ axis. The zero
             total energy contour is drawn with thick line and the domains of
             negative (dashed lines) and positive total energy (solid lines) are
             also shown.
             }
    \end{center}
\end{figure}

Let us first analyze the motion along the $z$-axis. The gravitational potential
in Eq.~\eqref{eq:Ring_Potential} simplifies significantly, as $r_{\!\perp}\equiv
0$ renders $k$ to be zero too, thus the elliptic integral, $K(0)=\frac{\pi}{2}$,
is only a constant in this expression. The potential is then
\begin{equation} \label{eq:Ring_Potential_AlongZAxis}
    V(z) = - \frac{G M}{p} = - \frac{GM}{\sqrt{R^{2} + z^{2}}},
\end{equation}
as expected. Because of this reduced effective dimensionality, we can easily
create a contour plot, see Figure~\ref{fig:Ring_PhasePlot_MotionAlongZAxis}, of
the total mechanical energy
\begin{equation}
    H(z, p_{z})
    = \frac{1}{2m}\, p_{z}^{2} + U(z)
    = \frac{1}{2m}\, p_{z}^{2} - \frac{G M m}{\sqrt{R^{2} + z^{2}}}
\end{equation}
over the phase space. The energy level curves $H(z, p_{z})= E$, where partition
the phase space $(z, p_{z})$ into three, qualitatively different types of
orbits. In Figure~\ref{fig:Ring_PhasePlot_MotionAlongZAxis} the dashed lines
correspond to total energy being negative, $H(z, p_{z}) < 0$, and these
trajectories are closed, i.e., in these cases the motion of the particle is
periodic. The separatrix, $H(z, p_{z}) = 0$, the curve separating the bounded
and unbounded trajectories, is emphasized in the same figure with thick solid
line. While this motion is not limited in coordinate space, the particle reaches
$z \rightarrow \pm \infty$, however, its speed is zero at infinity. Trajectories
with positive total energy are also unbounded, but the particle reaches infinity
with a finite velocity.
Here we mention that this Hamiltonian --after re-scaling-- coincides with that
of for the century-old MacMillan problem,\cite{MacMillan1911} a special
three-body problem.

\begin{figure}[h!]
    \begin{center}
    \includegraphics[width=80mm]{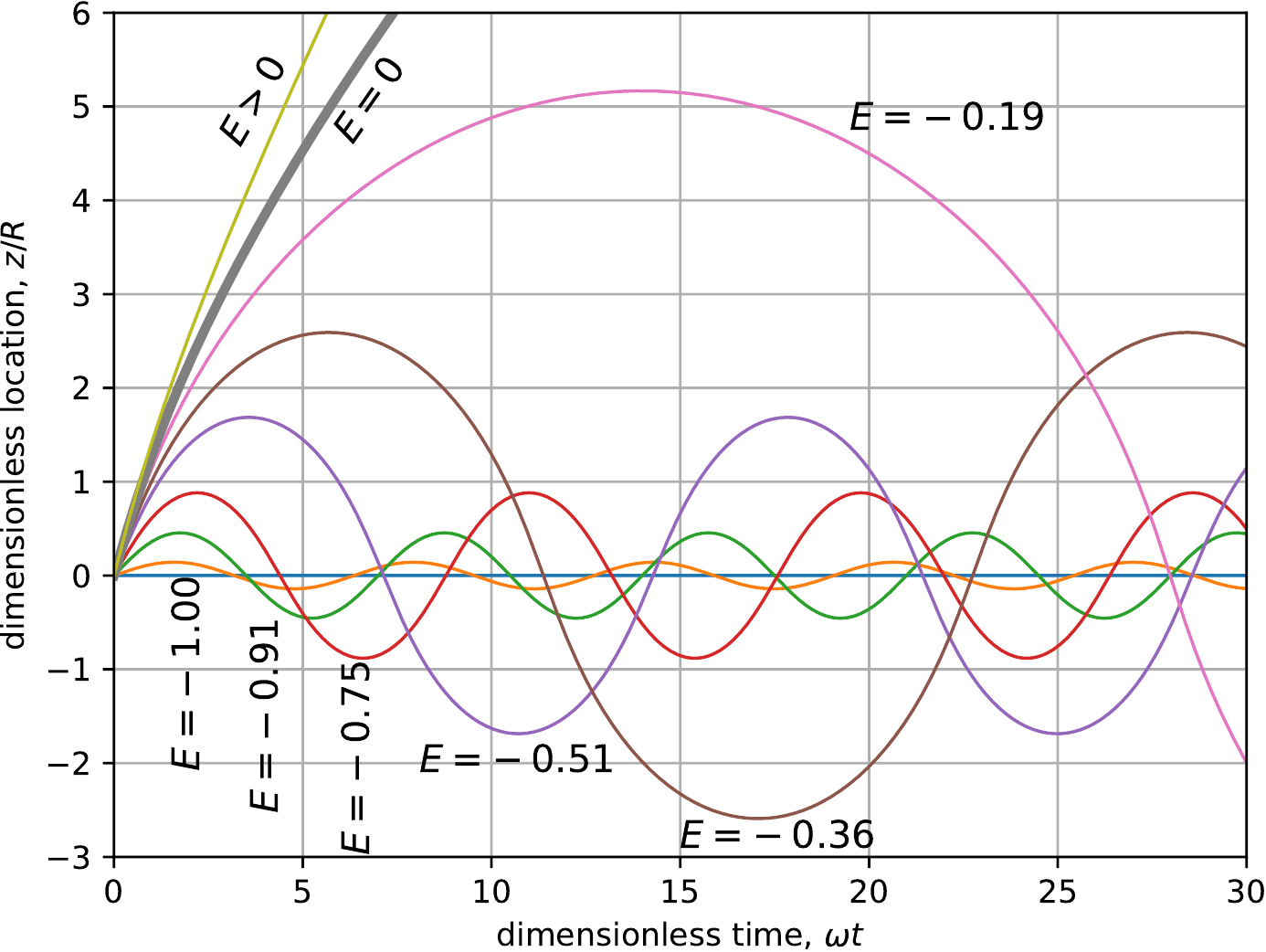}
    \caption{\label{fig:Ring_Trajectories_MotionAlongZAxis}
             (Color online) The time evolution of the $z$ coordinate of a unit
             mass particle along the $z$-axis. The particle initially is located
             at the center of the ring ($z=0$) but its initial momentum is
             varied so that its total energy is negative, zero or positive. The
             energy is given in dimensionless value, $E = H(z, p_{z})/(m R^{2}
             \omega^{2})$, where $\omega$ is the natural angular frequency
             defined as $\omega^{2} = GM/R^{3}$.
             }
    \end{center}
\end{figure}

In Figure~\ref{fig:Ring_Trajectories_MotionAlongZAxis} we plot the instantaneous
deflection, $z(t)$, of a particle moving along the $z$ axis. The corresponding
total energy is also given for all paths. As one expects, for lower, negative
energies, the amplitude is small and the motion resembles harmonic motion.
However with increasing energies the $z(t)$ curve deviates from a sinusoidal
undulation as a function of time and its shape morphs into more of half-circles,
see for example $E=-0.19$, and the period increases rapidly. At $E=0$ the
function $z(t)$ loses its periodicity. In the $z$--$t$ graph this path separates
the periodic paths from the diverging paths.

It is worth investigating how the period depends on or varies with the total
energy. After a significant amount of algebraic manipulations one may express
the period of the closed orbits as
\begin{equation} \label{eq:Ring_GeneralExpressionForPeriod}
    T
    =
    \frac{4}{\omega}
    \int_{0}^{1}
        {\frac{1}{(1-2k^{2} v^{2})^{2} \sqrt{(1-v^{2})(1-k^{2} v^{2})}} \,dv}
\end{equation}
where $\omega^{2} =GM/R^{3}$, the modulus $k^{2}=\frac{1}{2}(1-\abs{\epsilon})$,
and parameter $\epsilon = E/(m R^{2} \omega^{2})$ is the dimensionless energy of
the moving particle. The integral cannot be expressed in terms of elementary
functions, however, its structure suggests that a closed form can be given in
terms of complete elliptic integrals. We omit analysing this closed form, as
the main characteristic features of the period can be deduced from the integrand
directly\footnote{For the sake of completeness we provide here the exact
expression for the period
\unexpanded{\begin{equation*}
    T
    =
    \frac{4}{\omega}\,
    \frac{1}{2(1-2k^{2})}\,
    \Bigl \lbrack 2E(k^{2}) - K(k^{2}) + \Pi(2 k^{2} \vert k^{2}) \Bigr \rbrack
\end{equation*}}
in terms of the complete elliptic integrals of the first, second and third kind,
$K$, $E$, and $\Pi$, respectively. Their definitions can be found in
\citet{Byrd1954}. The power series expansion of this exact expression, up to
second order, agrees with that of given in the text, $T \approx
\tfrac{2\pi}{\omega} (1+\tfrac{9}{4}k^{2})$, but derived from a hand-waving
argument.}.

First, the integrand has a singularity at $v=1$, due to the $(1-v^{2})^{-1/2}$
term, irrespective of the value of $k$. However, this singularity on its own is
quite mild and, provided $k < \frac{1}{2}$, the other two terms remain finite
and the integrand remains integrable.

Second, small amplitude motion along the $z$ axis around the center of the ring
corresponds to $\epsilon \approx -1$, thus $k \approx 0$, and can be thought of
as a harmonic oscillation around the center. For small $k$ values one may expand
the integrand in Taylor series, and integrate term-by-term
\begin{equation*}
    T
    \approx
    \frac{4}{\omega}
    \int_{0}^{1}
        {\left \lbrack
         \frac{1}{\sqrt{1 - v^{2}}} + \frac{9 v^{2}}{2 \sqrt{1-v^{2}}} \, k^{2}
         \right \rbrack dv
        }
    =
    \frac{2\pi}{\omega}
    \!\left \lbrack 1 + \frac{9}{4} k^{2} \right \rbrack
\end{equation*}
The leading order is the classical result, $T = 2\pi/\omega$. This result also
gives the first correction as the energy increases. Although further terms of
the Taylor expansion have been omitted, it seems reasonable that all these
terms contain even powers of $k$, thus all terms will be positive as energy
increases towards zero.

Third, as the energy approaches zero, or alternatively as $k \rightarrow
\frac{1}{2}$, two terms become singular at $v=1$ in the integrand of
Eq.~\eqref{eq:Ring_GeneralExpressionForPeriod}. In the following we examine the
behavior of the integrand for $k = \frac{1}{2} - \delta k$, where $\delta k \ll
1$. The third term, $1/\sqrt{1-k^{2} v^{2}}$ is not singular, it remains finite
for all possible $k$ and $v$ values, thus we omit it as it can be thought of a
weight-function. Keeping this term would not change the $\epsilon$-dependence of
the period, but would only contribute a constant pre-factor. The remaining
integrand suggest a change of variable, $v = \sin{\!(x)}$, where $x$ runs over
$(0, \frac{\pi}{2})$. In this new variable the integration can be carried out
analytically if $k \ne \frac{1}{2}$.
\begin{equation*}
    T
    \approx
    \frac{4}{\omega}\!
    \int_{0}^{\pi/2}
        {\hspace*{-4mm}\frac{dx}{\lbrack 1 - 2 k^{2} \sin{\!(x)} \rbrack^{2}}}
    =
    \frac{4\pi}{\omega} \frac{1-k^{2}}{(1-2k^{2})^{\frac{3}{2}}}
    \approx
    \frac{2\pi}{\omega} \!\abs{\epsilon}^{-\frac{3}{2}}
\end{equation*}
\begin{figure}[t!]
    \begin{center}
    \includegraphics[width=80mm]{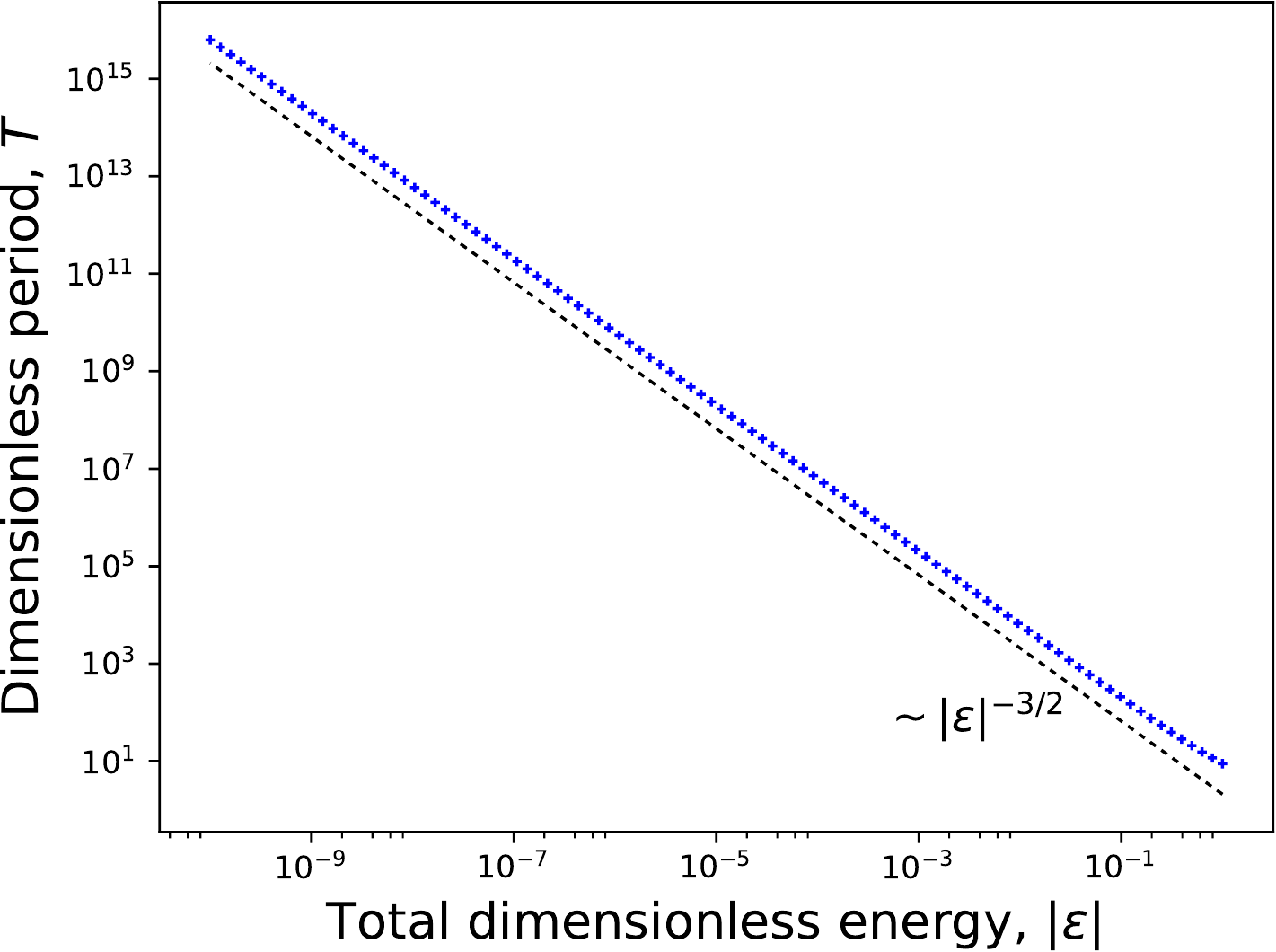}
    \caption{\label{fig:Ring_Period_MotionAlongZAxis}
             (Color online) The period, calculated numerically from
             Eq.~\eqref{eq:Ring_GeneralExpressionForPeriod}, is shown for
             trajectories with different total energy, $-1 \le \epsilon \le 0$
             ($0 \le k \le \frac{1}{2}$) on a log-log plot. In order to plot the
             logarithm of the energy the absolute value of $\epsilon$ is taken,
             thus higher energies are towards the left. The blue crosses
             represent the numerical calculation, while the black dashed line
             is only included to guide the eye and represents the asymptotic
             behavior of $T$ as $\epsilon \rightarrow 0^{-}$. It might be
             somewhat surprising how well this asymptote captures the energy
             dependence of $T$, and perhaps the only slight deviation is
             observed as $\abs{\epsilon} \rightarrow 1$.
             }
    \end{center}
\end{figure}
\noindent
Figure~\ref{fig:Ring_Period_MotionAlongZAxis} depicts the period, $T$ as a
function of the absolute value of the dimensionless energy, $\abs{\epsilon}$.
The derivation above suggested a power-law dependence, thus a log-log scale plot
is chosen with the absolute value of the energy as abscissa. Therefore, paths
with lower energy are towards the right of the plot, while the period of
trajectories with energy close to zero are towards the left of the abscissa.
The blue crosses represent the numerical values of periods as calculated from
Eq.~\ref{eq:Ring_GeneralExpressionForPeriod} via quadrature, while the thin
dashed line represent the power-law, $\sim \abs{\epsilon}^{-3/2}$, derived from
our approximation. The two ``curves'' seem to be parallel, indicating that there
is a constant multiplicative factor missing from our approximating power-law.
This is indeed the result of us completely neglecting one, non-singular term in
the integrand. Although we may have expected this approximation to be valid
only as $\epsilon \rightarrow 0^{-}$, it proves to be a surprisingly good
approximation in the entire interval examined. One might think that for even
lower energies the deviation between the numerical values and the power-law
would increase, as the slight curve appears in the blue crosses towards $\abs{
\epsilon} \approx 10^{-1}$, however, we should keep in mind that the energy
of a particle moving along the $z$ axis cannot be arbitrarily small, because the
gravitational potential energy has a local maxima at $r_{\!\perp} = 0$, namely
$U(r_{\!\perp})=-GMm/R$ as seen in Fig.~\ref{fig:Ring_PotentialAtFixedHeight}.
Since the kinetic energy is always a non-negative quantity the minimal value of
the total energy, $E$, corresponds to the minimal value of the gravitational
potential along the $z$ axis, i.e., $\text{min}(E) = -GMm/R$. Thus $\text{min}(
\epsilon) = \text{min}(E)/(mR^{2} \omega^{2}) = -1$.

\subsection{Motion in the $[xy]$-plane}

If a particle is initially located in the equatorial plane of the ring and its
initial velocity also falls into this plane then the entire trajectory of the
particle will remain in the $[xy]$ plane. Furthermore, the cylindrical symmetry
guarantees that the force at any point $P=(x, y, 0)$ points towards the origin,
thus the motion in this plane is a central force problem. Consequently there is
a supplementary conserved quantity on top of total mechanical energy, namely,
the angular momentum, $\bm{L}$ with magnitude $\abs{\bm{L}} = L$.

A particle does have two degrees of freedom moving within a plane. However, the
conserved quantities help us reducing the number of equations of motion to one
(see \S{14} in \citet{Landau1982})
\begin{equation*}
    m \ddot{r}_{\!\perp}
    =
    \frac{{L}^{2}}{m r_{\!\perp}^{3}}
    -\frac{d}{dr_{\!\perp}}
    U \!\left ( r_{\!\perp},\vartheta = \frac{\pi}{2} \right )
    =
    -\frac{d}{dr_{\!\perp}} U_{\mathrm{eff}}(r_{\!\perp}, L).
\end{equation*}
\begin{figure}[b!]
    \begin{center}
    \includegraphics[width=80mm]{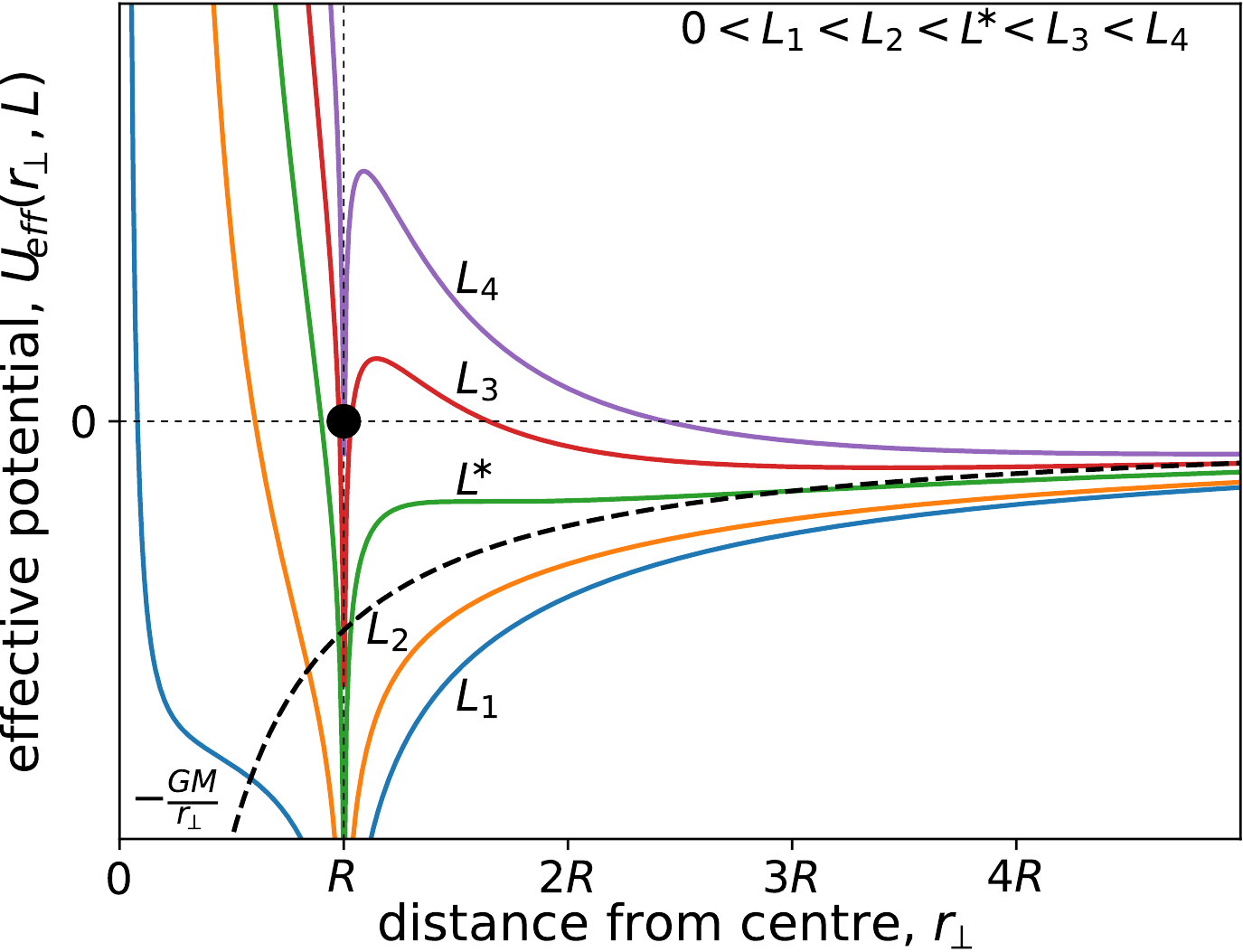}
    \caption{\label{fig:Ring_EffectivePotential_xy}
             (Color online) The effective potential, $U_{\mathrm{eff}}(
             r_{\!\perp}, L)$, is plotted as a function of $r_{\!\perp}$ for
             five values of angular momenta. As guidance, the gravitational
             potential of a point particle, with identical mass as the ring, is
             also depicted (dashed line).
             }
    \end{center}
\end{figure}
The first term can be written as a derivative with respect to $r_{\!\perp}$,
thus one might combine these two terms and interpret the equation of motion so
that the particle is under the influence of the {\emph{effective}} potential
\begin{equation} \label{eq:EffectivePotential_MotionEquatorialPlane}
    U_{\mathrm{eff}}(r_{\!\perp}, L)
    =
    \frac{{L}^{2}}{2m r_{\!\perp}^{2}}
    + U \!\left ( r_{\!\perp}, \vartheta=\frac{\pi}{2} \right ).
\end{equation}

Figure~\ref{fig:Ring_EffectivePotential_xy} shows the effective potential for
four values of $L$ and unit mass, $m=1$. There are several features of the
graphs which are worth some analysis. First, the singularity at $r_{\!\perp}=R$
is due to the inherent singularity of the gravitational potential itself and it
is not the consequence of the specific geometry. Second, irrespectively of
angular momentum the effective potential is strictly decreasing for $r_{\!\perp}
< R$. This behavior gives further support to the fact that there cannot be a
stable trajectory ``inside'' the ring; for increasing angular momentum the
effective potential is stepper and steeper towards the ring. Even the $L=0$
case, i.e., the motion parallel to the $z$-axis, is not an exception. It is
unstable against perturbations, since the effective potential develops a local
maximum for $r_{\!\perp}=0$ and $L=0$ and not a minimum. Third, in the case of
$R < r_{\!\perp}$, one may see that for a yet undetermined, threshold value
$L^{\ast}$, a shallow but wide minimum develops in the effective potential. This
feature indicates that stable, periodic orbits may exist further away from the
ring.
\begin{figure}[H]
    \begin{center}
    \includegraphics[width=80mm]{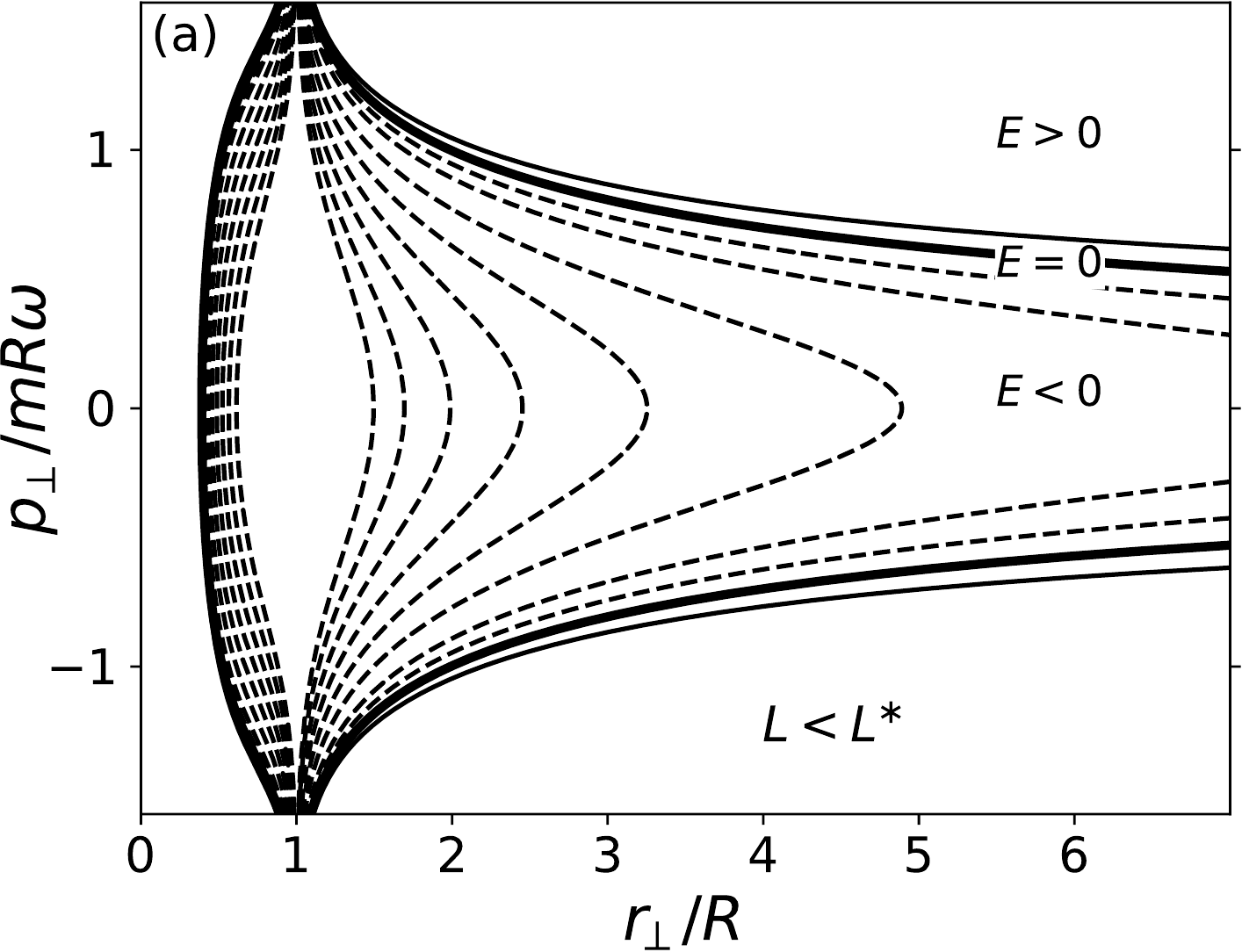}
    \includegraphics[width=80mm]{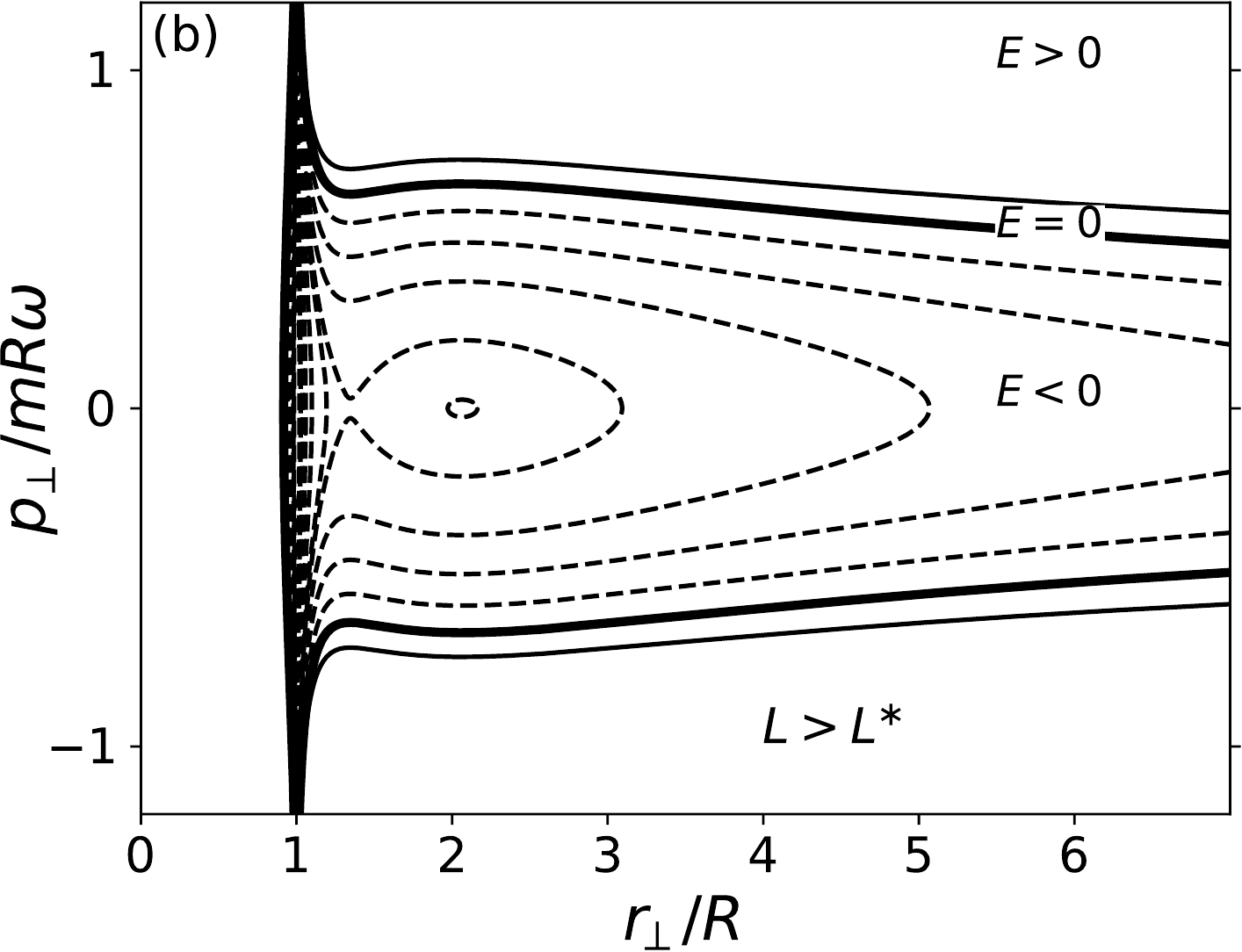}
    \includegraphics[width=80mm]{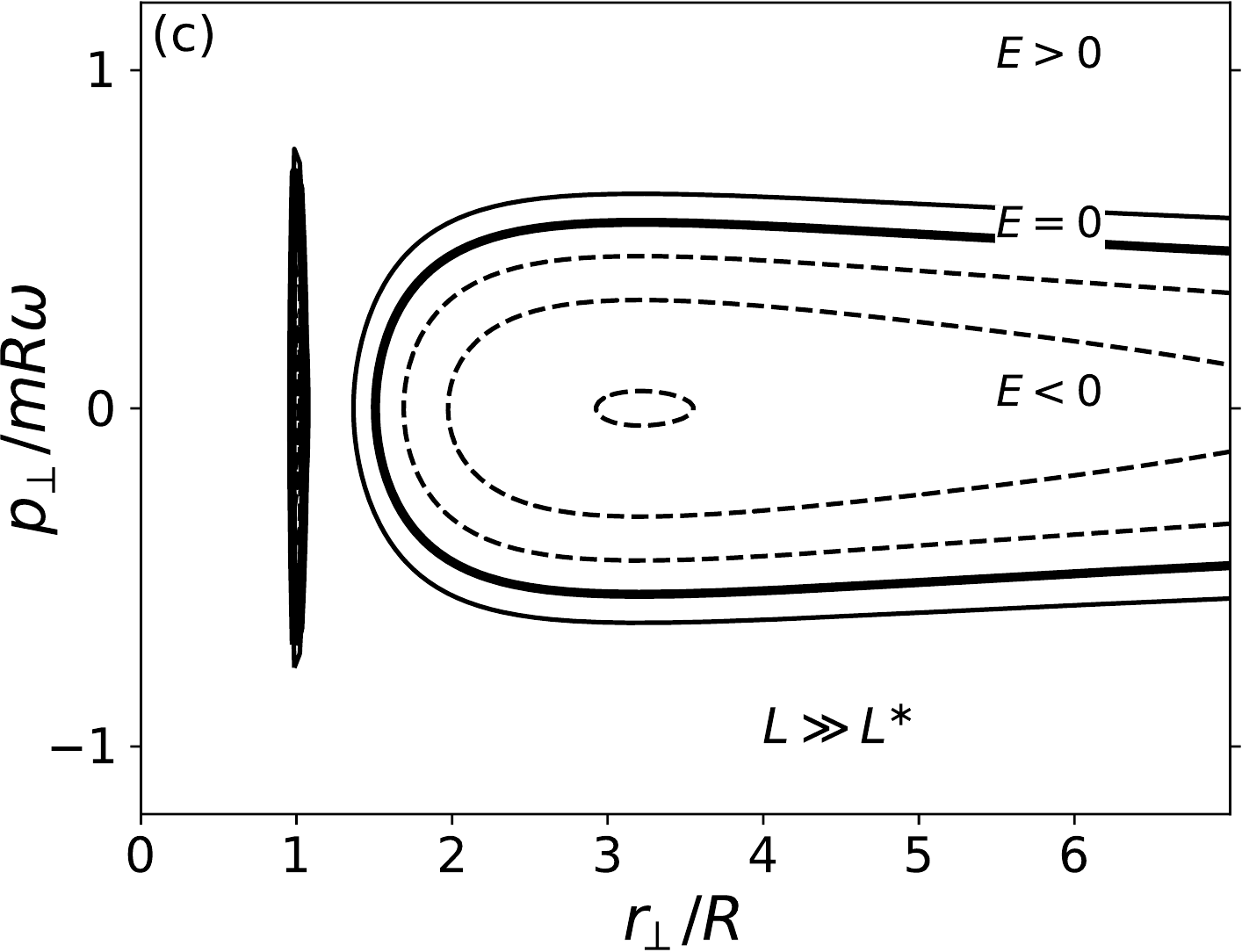}
    \caption{\label{fig:Ring_EnergyContours_xy_OverPhaseSpace}
             Contour plot of the total mechanical energy, $H(r_{\!\perp},
             p_{\!\perp})$, over the phase space for a particle moving in the
             equatorial plane for three values of the total angular momentum,
             $L< L^{\ast}$, $L \approx L^{\ast}$, and $L \gg L^{\ast}$. The zero
             total energy contour is drawn with thick, solid line and the
             domains of negative (dashed lines) and positive total energy (solid
             lines) are also shown. The birth of a small dip in the energy is
             clearly visible for increasing angular momentum. Also notice that
             in the $r_{\!\perp}<R$ region all energy contour lines, both
             negative and positive are accumulating around the singularity.
             Finally the asymmetry of the energy minimum is also observable,
             especially for higher energies.
             }
    \end{center}
\end{figure}
Thus a particle with the corresponding energy and appropriately chosen initial
location can be in a stable, dynamical equilibrium in real space. For further
increasing the angular momentum (subfigure (c)) the potential well grows in size
(more and more closed lines appear in the plot) and also moves away from ring as
expected.

It seems worthwhile to find out the critical angular momentum, $L^{\ast}$, at
which the dip appears in the effective potential.
\begin{figure}[b!]
    \begin{center}
    \includegraphics[width=80mm]{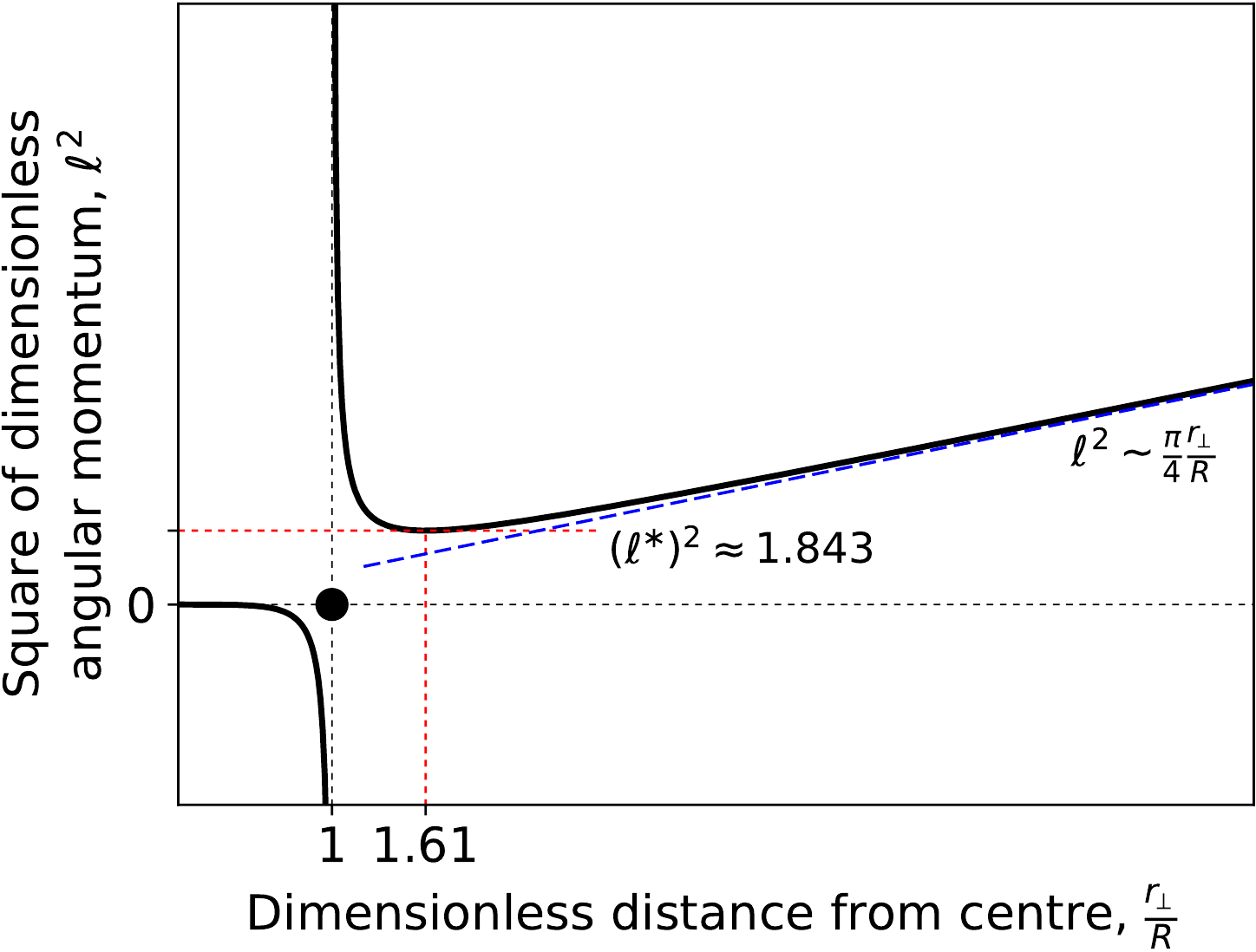}
    \caption{\label{fig:Ring_MinimalAngularMomentum_xy}
             (Color online) The square of dimensionless angular momentum,
             $\ell^{2}$, corresponding to the stationary point of the effective
             potential is plotted as a function of the dimensionless distance,
             $r_{\!\perp}/ R$. Consequently, the $r_{\!\perp} < R$ region does
             not support a physically acceptable solution, since $\ell^{2}$ is
             negative in this interval. On the outer side of the ring one needs
             a critical value for $\ell^{2}$ so that the effective potential has
             a stationary point, or in other words, the particle does not
             collide with the ring.
             }
    \end{center}
\end{figure}
In order to capture this peculiar behavior, i.e., the appearance of a critical
angular momentum value, we may calculate the derivative of the effective
potential in Eq.~\eqref{eq:EffectivePotential_MotionEquatorialPlane}, with
respect to $r_{\!\perp}$ and equate it with zero. Doing so and expressing
$L^{2}$ we arrive at
\begin{equation} \label{eq:DetermineCriticalAngularMomentum}
    \ell^{2}
    =
    - \left ( \frac{x}{2} \right )^{\!2} \frac{1}{1+x}
    \left \lbrack \frac{1+x}{1-x} \, E(k) - K(k) \right \rbrack,
\end{equation}
where $\ell^{2} = L^{2}/ \ell_{0}^{2}$, with $\ell_{0}^{2} = 4GMm^{2}R/\pi$, is
the square of the dimensionless angular momentum, and $x=r_{\!\perp}/R$ is the
dimensionless distance in $[xy]$-plane. Naturally the left-hand-side of this
equation needs to be non-negative, therefore this equation has physical
relevance only for those $x$ values for which the right-hand-side is also
non-negative, and then this equation provides the pair ($x$, $\ell^{2}$), or
equivalently ($r_{\!\perp}$, $L^{2}$), for which the effective potential has
local extremum.

The right-hand-side of Eq.~\eqref{eq:DetermineCriticalAngularMomentum} is
plotted as a function of $x$ (the modulus, $k$, is also a function of $x$).
Inside the ring ($x<1$) there is no physical solution, as $\ell^{2}$ is
negative. Consequently there is no path inside the ring which could be stable;
a moving particle is attracted to the ring and inevitably collides with it.
However, outside of the ring, we observe that there is either no physical
solution, or there are two. The curve defines a minimal distance $x^{\ast} \approx
1.61$, and minimal angular momentum, $\ell^{\ast} \approx \sqrt{1.843} \approx
1.36$, at which a local minimum starts developing in the effective potential. We
need to distinguish two segments of the curve in the $x>1$ region depending on
the relative values of $x$ and $x^{\ast}$. The interval $1 \le x \lesssim 1.61$
corresponds to the local maximum of the effective potential (see
Figure~\ref{fig:Ring_EffectivePotential_xy}) and thus represents unstable
orbits. However, the $(x, \ell^{2})$ pairs in the $1.61
< x$ domain all correspond to the local minimum of the effective potential and
define stable orbits in the gravitational field of the ring provided the
motion is confined to the $[xy]$ plane.

The Taylor expansion of the complete elliptic functions, $E(k)$ and $K(k)$,
around $k \approx 0$ would also provide the asymptotic behavior of the curve in
Fig.~\ref{fig:Ring_MinimalAngularMomentum_xy}, i.e., $\ell^{2} \sim
\frac{\pi}{4} x$. Here, we recovered the classical result for circular orbits
around a point-like attractive center, i.e., $L_{\text{cl}}^{2} = GMm^{2} r$,
where $r$ is the distance from the center. Using the unit of $\ell_{0}^{2}$
again, the dimensionless square of the classical angular momentum is
$\ell^{2}_{\text{cl}} = L_{\text{cl}}^{2}/\ell_{0}^{2} = \frac{\pi}{4} r/R$.
Figure~\ref{fig:Ring_MinimalAngularMomentum_xy} shows this classical result as
dashed blue line and it is the asymptote for $\ell^{2} \sim \ell^{2}_{\text{cl}}$
as $r_{\perp} \gg R$.

Finally we demonstrate that stable and periodic, circular orbits do exist in the
equatorial plane of the ring. Furthermore, our numerical experiments indicate
that these orbits are stable not only against in-plane perturbations but against
out-of-plane perturbations as well. In calculating the trajectory of a point
particle we rely on the explicit expression of the gravitational force given in
equation~\eqref{eq:GeneralMotionInTheFieldOfRing}.

\begin{figure}[t!]
    \begin{center}
    \includegraphics[width=85mm]{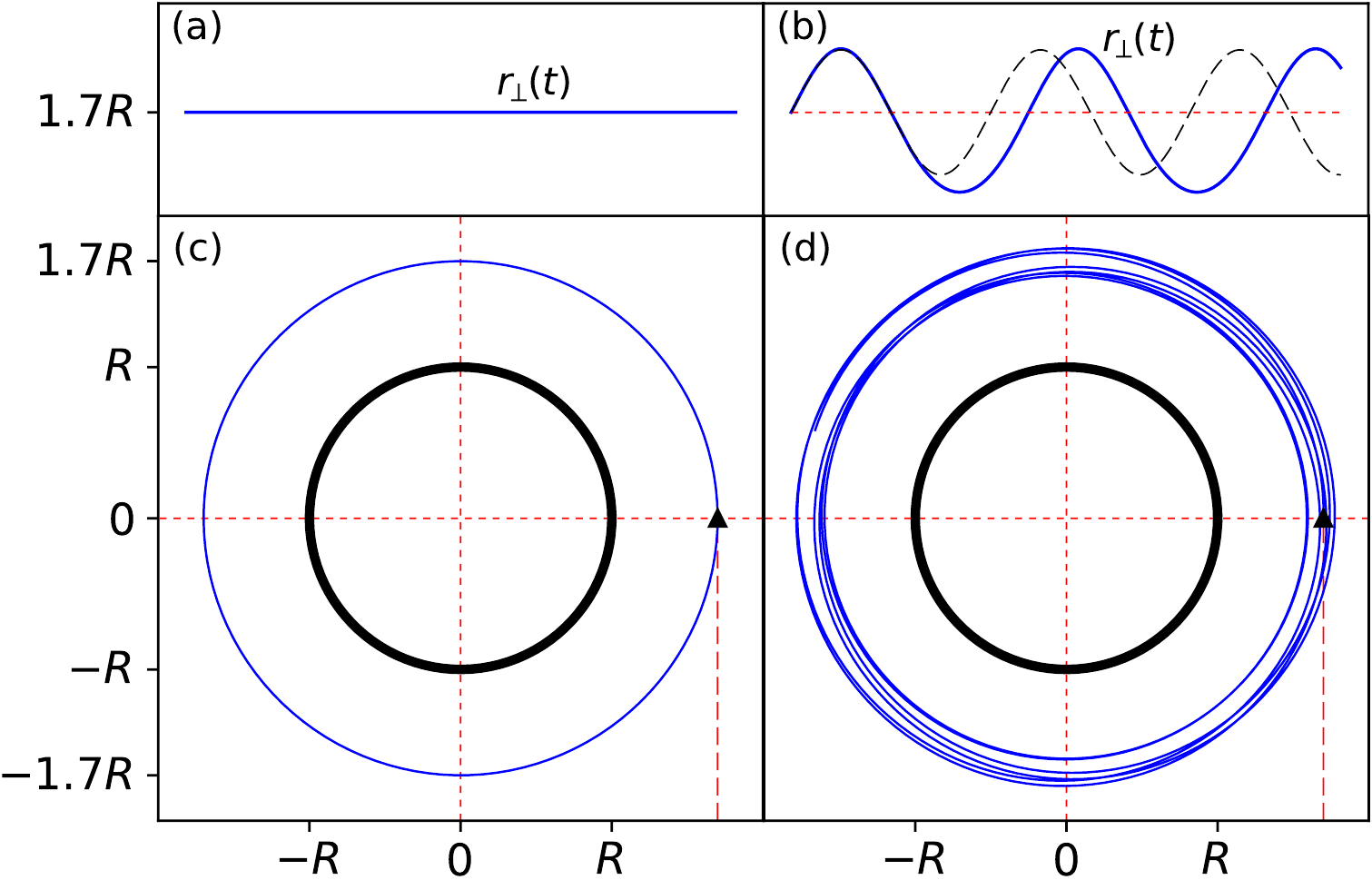}
    \caption{\label{fig:Ring_CircularAndPerturbedEquatorialOrbits}
             (Color online) The time evolution of two orbits are shown. The
             bottom row contains an overview of the $xy$ plane showing the ring
             (black solid line), the initial location at $r_{\!\perp}=1.7R$
             (black upright triangle) and the orbits (blue solid line). In
             subplot (c) the initial velocity points in the $\bm{e}_{y}$
             direction and its magnitude was chosen such that the orbit is
             circular and thus periodic. In subplot (d) the initial velocity of
             (c) was slightly perturbed in the $\bm{e}_{x}$ direction, however,
             the magnitude of this perturbation is approximately one fiftieth of
             the $y$ component. In subplots (a) and (b) the time evolution of
             $r_{\!\perp}(t)$ is plotted. The actual timescale is not shown as
             that is immaterial to our analysis here. The scale for the
             deviation in (a) and (b) is the same and it is chosen so that the
             maximum deviation --approximately a hundredth of $1.7R$ value-- in
             the perturbed case is clearly visible.
             }
    \end{center}
\end{figure}

\begin{figure}[t!]
    \begin{center}
    \includegraphics[width=85mm]{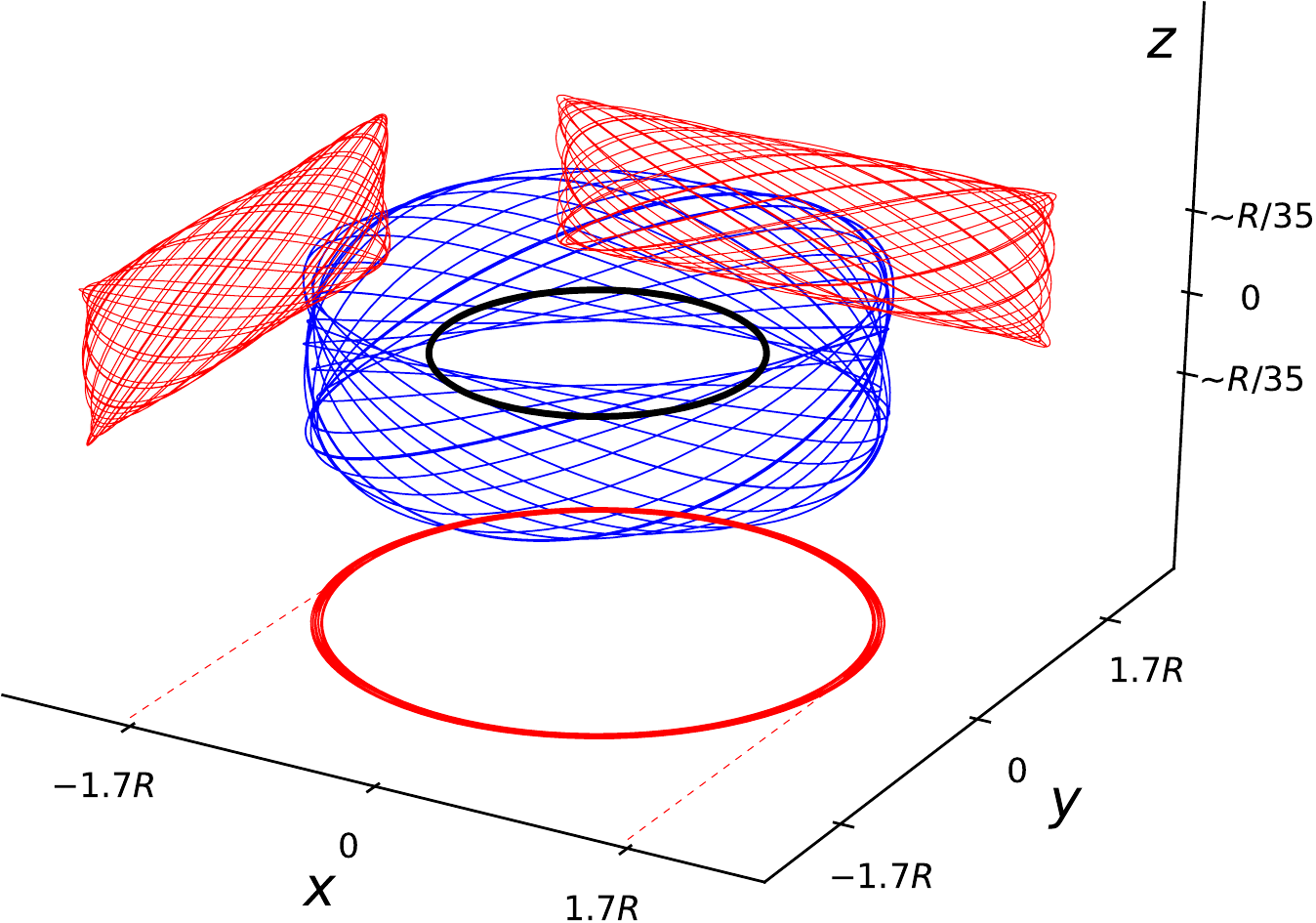}
    \caption{\label{fig:Figure_Ring_CircularAndPerturbedOrbits_XZDirection}
             (Color online) The time evolution of an orbit and its
             two-dimensional projections are shown. The solid black circle in
             the $z=0$ plane represents the massive ring. The initial velocity
             used for Fig.~\ref{fig:Ring_CircularAndPerturbedEquatorialOrbits}(a)
             is perturbed both in the $x$ and $z$ directions, while the initial
             location, $(1.7R, 0, 0)$ is kept unchanged. It is clear that while
             the orbit is not periodic any more, it is stable, and confined in
             each direction.
             }
    \end{center}
\end{figure}

Figures~\ref{fig:Ring_CircularAndPerturbedEquatorialOrbits} and
\ref{fig:Figure_Ring_CircularAndPerturbedOrbits_XZDirection} show
trajectories of particles in the gravitational field of a massive ring. The
initial conditions were chosen so that the path is periodic (Fig.
\ref{fig:Ring_CircularAndPerturbedEquatorialOrbits}c), slightly perturbed
in-plane (Fig.~\ref{fig:Ring_CircularAndPerturbedEquatorialOrbits}d) or
perturbed both in-plane and out-of plane
(Fig.~\ref{fig:Figure_Ring_CircularAndPerturbedOrbits_XZDirection}).

Subplot (c) shows the ring (black inner circle) and the orbit (blue circle)
in the $[xy]$-plane, with the black triangle indicating the initial location,
$r_{\!\perp} = 1.7R$. The initial velocity is entirely in the $y$ direction and
its magnitude is chosen so that a circular path is formed. The time evolution of
the particle's distance from the center, $r_{\!\perp}(t)$ is shown above, in
subplot (a) as a function of time. It is apparent that it does not change more
than the numerical floating-point precision. The orbit on the right, in subplot
(d), is started as in (c) but its initial velocity is perturbed in the $x$
direction. It is clearly visible in
Fig.~\ref{fig:Ring_CircularAndPerturbedEquatorialOrbits}(d) that the orbit does
not close any more, but it forms a rosetta, confined between two concentric
circles. The stability --at least against perturbation within the equatorial
plane-- is indicated by subplot (b) depicting a seemingly periodic variation of
$r_{\!\perp}(t)$ around the exact circular path with radius $r_{\!\perp}=1.7R$.
It is easily interpreted based on the effective potential in
Fig~\ref{fig:Ring_EffectivePotential_xy}. As the effective potential develops a
dip in the $r \gtrsim 1.61$ region a slight perturbation should drive the
particle into an oscillation around the dip. As the effective potential is
asymmetric around the minimum, see
Fig.~\ref{fig:Ring_EnergyContours_xy_OverPhaseSpace}, the oscillation around the
minimum is anharmonic. That is indeed what we observe in
Fig.~\ref{fig:Ring_CircularAndPerturbedEquatorialOrbits}(b). The black dashed
line is a sinusoidal fit to the very first half-cycle. As time progresses the
fitted sinusoidal variation and the actual oscillation deviate from each other,
demonstrating an anharmonic oscillation. Similar motion is present in
astronomical systems, when the relative distance between two objects, e.g.,
planet and its moon, varies slightly with time.

Similar calculations can be made by modifying the perturbing velocity component,
for example in the $z$-direction as well. Here, without detailed analysis, we
include only one figure,
Figure~\ref{fig:Figure_Ring_CircularAndPerturbedOrbits_XZDirection}, showing the
trajectory of a particle (blue line) in the gravitational field of a massive
ring (black circle). The particle is released with nearly identical initial
conditions as in Fig.~\ref{fig:Ring_CircularAndPerturbedEquatorialOrbits}(d),
but a small velocity component in the $z$-direction was also added to its
initial velocity. The projections of its trajectory is also shown in all three
planes drawn with red lines. The $[xy]$ projection is similar to that of in
Fig.~\ref{fig:Ring_CircularAndPerturbedEquatorialOrbits}(d), i.e., the path
remains between two concentric circles,  while the other two projections
confirm that the path remains within well-defined limits (boxes).
In other words, the inhabitants of Ringworld could also invest in launching
satellites which are whizzing around their peculiar planet and provide the
Ringlings with GPS. The general relativistic calculation for the temporal shift
to their clocks is, however, beyond the scope of this article.

Finally we mention that the non-central gravitational field allows unusual,
{\emph{periodic}} orbits which are absent in central symmetric fields. It is a
well-known and often taught result of classical mechanics (Bertrand's problem)
that central force fields in which all bounded paths are closed are that of the
harmonic oscillator ($\bm{F} = -\alpha \bm{r}$) and of the spherically symmetric
gravitational attraction ($\bm{F} \sim - \alpha \bm{r}/\abs{\bm{r}}^{-3}$),
where $\alpha$ is a positive constant.\cite{Bertrand1873, Goldstein1980,
Arnold1989, Grandati2008} The non-central field of a massive ring analyzed in
this work, provides a relatively simple case where one can construct periodic
bounded orbits, e.g., in the shape of $\infty$ in a meridional plane, where the
crossing of $\infty$ coincides with the center of the ring. In this work,
however, we wanted to focus on the geometric structure of the field, the
constraints, e.g., what physical quantity is conserved, and have only touched on
the analysis of the dynamics of a point particle in this field.

\section{\label{sec:Conclusion} Conclusion}

In this paper we have analyzed the gravitational field of a massive, filamentary
ring and investigated some of the possible orbits of a point-like, massive
particle moving in the field of that ring. The choice of a ring was motivated
by the high symmetry of this configuration; it possesses continuous rotational
symmetry around the $z$ axis, reflection with respect to the plane of the ring.

We provided an analytic, closed-form expression for the gravitational potential.
One might approach this problem by comprising the entire mass of the ring into
its geometric center, and thus calculate at any point in space what potential
this point-like object created. However, we demonstrated that the exact
gravitational potential and its gradient, thus the gravitational force-field, is
not central. Depending on the location of the ``test mass'' the gravitational
attraction exerted on it points at different locations.

Another consequence of the reduced symmetry, or in other words, the non-central
force-field is that the angular momentum, in general, is not conserved. However,
the continuous rotational symmetry around the fixed $z$ axis guarantees that the
$z$ component of the angular momentum, $L_{z}$, is indeed a conserved quantity
irrespectively of the motion. Furthermore the high symmetry of this system also
ensures that there are special classes of motions which are restricted to
invariant subspaces, i.e., motion along the $z$-axis, motion in the equatorial
$[xy]$-plane or motion within any meridional plane. This means that the
trajectory of a particle remains in a plane if its initial velocity also
parallel to this plane. In this special cases the motion is integrable, as the
particle has only one-degree of freedom.

We have also shown by analyzing the equatorial motion that inside the ring all
trajectories are unstable and eventually the particle collides with the ring.
However, outside of the ring periodic orbits are possible, but only if the
angular momentum is above a threshold value, $L^{\ast}$. In that case the
circular orbit is stable against small perturbations, even if the perturbation
is out of plane in contrast to what might naively be expected from symmetry.

Solving the equations of motion numerically revealed that peculiar, and
{\emph{periodic}} orbits with a shape resembling $\infty$ exist in the
meridional planes. Such orbits would not be present in a central force-field,
since they would inevitably lead to a collision with the central object.

A more detailed investigation of the dynamics of a point particle in the field
of a ring would be warranted as it seems to contain unusual orbits. Furthermore
this system --due to its high symmetry-- seems to be a great candidate for
analyzing the integrability of the dynamics.

\begin{acknowledgments}
    D.~S. acknowledges the financial support from The Dodd--Walls Centre for
    Photonic and Quantum Technologies.
\end{acknowledgments}

\end{document}